\documentclass[prb,10pt,english, twocolumn,aps,showpacs,superscriptaddress,longbibliography]{revtex4-2}

\usepackage{mathtools}
\usepackage{graphicx} 
\usepackage{xcolor}
\usepackage{adjustbox}
\usepackage[colorlinks = true,
            linkcolor = blue,
            urlcolor  = blue,
            citecolor = blue,
            anchorcolor = blue]{hyperref}
\usepackage[normalem]{ulem}
\usepackage{physics}
\usepackage{appendix}
\usepackage{footnote}
\usepackage{orcidlink}
\usepackage{mathrsfs}
\usepackage{booktabs}

\usepackage[draft,inline,nomargin]{fixme}
\fxsetup{inlineface=\mdseries} 
\definecolor{mggreen}{HTML}{1E6B42}
\FXRegisterAuthor{mg}{amg}{\color{mggreen}MG}
\makeatletter
\renewcommand*\FXLayoutInline[3]{%
  {\@fxuseface{inline}\ignorespaces#3: #2}}
\makeatother
\usepackage{tikz}
\usetikzlibrary{patterns,arrows.meta}

\newcommand{\barb}[1]{{#1}}

\newcommand{\pcsadd}{Center for Theoretical Physics of Complex Systems, Institute for Basic Science (IBS), Daejeon 34126, Republic of Korea}
\newcommand{\ctiqs}{Center for Trapped Ion Quantum Science, Institute for Basic Science, Daejeon 34126, Republic of Korea}
\newcommand{\ustadd}{Basic Science Program, Korea University of Science and Technology (UST), Daejeon 34113, Republic of Korea}
\newcommand{\nziasadd}{Centre for Theoretical Chemistry and Physics, The New Zealand Institute for Advanced Study (NZIAS), 
Massey University Albany, Auckland 0745, New Zealand}
\newcommand{\icn}{Instituto de Ciencias Nucleares, Universidad Nacional Aut\'onoma de M\'exico, Apdo. Postal 70-543, C.P. 04510  Cd. Mx., M\'exico}

\begin{document}

\title{Prethermalization of entanglement entropy in the mixed-field Ising model}

\author{Miguel Gonzalez\,\orcidlink{0009-0004-0112-5988}}
    \email{miguel.gonzalez@correo.nucleares.unam.mx}
    \affiliation{\icn}
    \affiliation{\pcsadd}

\author{Budhaditya Bhattacharjee\,\orcidlink{0000-0003-1982-1346}}
    \email{budhaditya.bhattacharjee@uni.lu}
    \affiliation{\pcsadd}
    \affiliation{Department of Physics and Materials Science, University of Luxembourg, L-1511 Luxembourg, Luxembourg}

\author{Alexei Andreanov\,\orcidlink{0000-0002-3033-0452}}
    \email{aalexei@ibs.re.kr}
    \affiliation{\pcsadd}
    \affiliation{\ctiqs}
    \affiliation{\ustadd}

\author{Barbara Dietz\orcidlink{0000-0002-8251-6531}}
    \email{bdietzp@gmail.com}
    \affiliation{\pcsadd}
    \affiliation{\ustadd}
    \affiliation{Max Planck Institute for the Physics of Complex Systems, N\"othnitzer Str. 38, 01187 Dresden, Germany}
    \affiliation{TU Dresden, Institute of Theoretical Physics, 01062 Dresden, Germany}

\author{Sergej Flach}
    \email{sergejflach@googlemail.com}
    \affiliation{\pcsadd}
    \affiliation{\ustadd}
    \affiliation{\ctiqs}
    \affiliation{\nziasadd}

\begin{abstract}

We study how weak integrability breaking affects the dynamics of the half-chain entanglement entropy of random product states in the mixed-field Ising model. The associated Hamiltonian depends on two parameters $g$ and $h$ determining the sizes of the longitudinal and transverse fields and has two integrable limits: the classical Ising chain $(h=0)$ and the transverse-field Ising model $(g=0)$. Weak perturbations of these limits give rise to the short-range network (SRN) and long-range network (LRN) regimes, respectively. We characterize the departure from integrable dynamics and the emergence of intermediate entanglement plateaus by melting times $T_M$ at prescribed entropy thresholds and analyze their dependence on the perturbation strength. For the SRN parameters examined away from $g=2$, we find $T_M\propto h^{-2}$ over the thresholds tested. At $g=2$, the relaxation splits into two stages: an initial departure from the integrable plateau with $T_M\propto h^{-1}$, followed by the melting of a well-defined intermediate prethermal plateau with $T_{M_2}\propto h^{-2}$. In the LRN regime, by contrast, the inverse of $T_M$ scales with exponents of $g$ close to either one or two, depending on the parameter values, the system size, and the entropy threshold, and no well-defined prethermal plateaus are observed. Our analysis distinguishes the melting of integrable dynamics from the subsequent decay of a prethermal plateau, which is observed in the SRN regime.
\end{abstract}

\maketitle

\section{Introduction}
\label{sec:intro}

Thermalization in isolated quantum many-body systems describes the relaxation of observables toward values consistent with statistical-mechanical equilibrium~\cite{DAlessio2016,Mori_2018}.
This relaxation, however, does not necessarily occur on a single characteristic timescale.
Different dynamical processes may relax on well-separated timescales, allowing the system to enter a long-lived intermediate regime before reaching its final thermal state~\cite{Mori_2018,Prethermalizationchaotic2019}.
Such a separation of timescales is a characteristic phenomenology of prethermalization~\cite{Prethermalization1}. Prethermal regimes have been identified in a broad range of settings, including near-integrable isolated quantum systems~\cite{Langen_2016,AduSmith_2013,science.1224953,Prethermalizationchaotic2019},
periodically driven many-body systems~\cite{PhysRevB.103.L140302,doi:10.1126/science.abg8102,KUWAHARA201696,PhysRevLett.131.130401}, and classical many-body systems~\cite{confinedchaos2025,PhysRevLett.122.010602,PhysRevLett.127.140602}.
The mechanisms responsible for the separation of timescales, and consequently the scaling of the associated relaxation times, depend on the structure of the system, the perturbation, the observable under consideration, and the initial state.

A common mechanism for the emergence of a prethermal regime is a weak breaking of an exact conservation law.
Consider a Hamiltonian $\hat H_0$ with a conserved operator $\hat Q$, such that $[\hat H_0,\hat Q]=0$.
We introduce a weak perturbation,

\begin{gather}
    \hat H(\lambda)=\hat H_0+\lambda\hat V,
    \qquad
    \lambda\ll1,
    \qquad
    [\hat V,\hat Q]\neq0,
\end{gather}

which breaks the conservation law.
For sufficiently small $\lambda$, however, its violation can occur on a parametrically long timescale, allowing the system to remain for an extended period in a quasi-stationary regime before its eventual thermalization~\cite{Mori_2018,Prethermalizationchaotic2019}.

In the particular case of an integrable quantum many-body system, the Hamiltonian $\hat H_0$ possesses an extensive set of mutually commuting conserved operators~\cite{Caux2011,Vidmar2016_GGE}.
The conserved operators $\hat Q_i$ satisfy

\begin{gather}
    [\hat H_0,\hat Q_i] = 0,
    \qquad
    [\hat Q_i,\hat Q_j] = 0.
\end{gather}

These conserved quantities strongly constrain the dynamics and, in general, prevent relaxation to the conventional thermal ensemble described by the Eigenstate Thermalization Hypothesis~\cite{sred1994chaos,deutsch2018eigenstate,DAlessio2016}.
An integrability-breaking perturbation $\lambda\hat V$ weakly violates these constraints whenever $[\hat V,\hat Q_i]\neq0$.
When $\lambda$ is small, the dynamics can therefore remain close to those of the integrable system over an extended time interval before the effects of integrability breaking become dominant.
In the entanglement dynamics studied below, we distinguish the departure from the integrable plateau from the decay of any distinct intermediate plateau. We call such an intermediate plateau a prethermal plateau.

The observable considered in this work is the dynamical bipartite entanglement entropy of random product states in the lattice basis.
For the same class of initial states, we compare independently the entanglement dynamics generated by the integrable Hamiltonian $\hat H_0$ and by the nearly integrable Hamiltonian $\hat H(\lambda)$ as a function of the integrability-breaking parameter $\lambda$.

To characterize the departure from integrable dynamics, we introduce a melting time $T_M$ and study its scaling with the integrability-breaking parameter $\lambda$.
We perform this analysis in the mixed-field Ising model (MFIM), which possesses two qualitatively distinct integrable limits.
Following the terminology introduced by analogy with the classical case~\cite{srnlrndefinitions2019,merab2022lyapunov},
we refer to the corresponding near-integrable regimes as short- and long-range networks.
In the short-range network (SRN) regime, we additionally identify a second melting time $T_{M_2}$ associated with the decay of an intermediate prethermal plateau.
Comparing these relaxation stages with the long-range network (LRN) dynamics allows us to distinguish the breakdown of integrable behavior from the melting of a prethermal state.

\section{Preliminaries}
\label{sec:model}

We consider the mixed-field Ising model ($\hbar=1$) with open boundary conditions (OBC), given by
\begin{gather}
    \hat{H} = -J\sum_{i=1}^{L-1}\hat{\sigma}^{z}_{i}\hat{\sigma}^{z}_{i+1} - g\sum_{i=1}^{L}\hat{\sigma}^{z}_{i} - h\sum_{i=1}^{L}\hat{\sigma}^{x}_{i} \, ,
    \label{eq:ising}
\end{gather}
with $\hat{\sigma}^{x,y,z}_{i}$ denoting the Pauli matrices.
For $J=0$, the system consists of $L$ independent spins.
We set $J=1$ throughout this work.
The two integrable limits correspond to $h=0$ and $g=0$, respectively, whereas nonintegrability has been proven for any $g \neq 0$ and $h \neq 0$~\cite{PhysRevB.109.035123}.

For $g=0$, the model reduces to the transverse-field Ising model (TFIM)~\cite{freefermion1970} {and $\hat H$ can be diagonalized using  Jordan--Wigner and Bogoliubov transformations,}
\[
    {
    \hat H_{\mathrm{TFIM}} =
    \sum_{\mu=1}^{L}\epsilon_\mu
    \left(\hat\eta_\mu^\dagger\hat\eta_\mu-\frac12\right).}
\]

{A complete set of mutually commuting conserved quantities is given by the quasiparticle occupations}

\begin{equation}
    \label{eq:LRNcharges}
    {
    \hat Q_\mu=\hat\eta_\mu^\dagger\hat\eta_\mu,
    \qquad \mu=1,\ldots,L.}
\end{equation}

{Their joint eigenvalues label the fermionic Fock eigenstates.}

For $h=0$, the model is classical and the conserved quantities are the local operators

\begin{gather}
    \hat{\mathcal{O}^{(k)}} = \hat{\sigma}^{k}_{z} \,.
    \label{eq:SRNcharges}
\end{gather}

The Hamiltonian of the model~\eqref{eq:ising} is exclusively composed of local operators for all parameter values. 
However, the conserved quantities behave qualitatively differently in the two integrable limits:
for $h = 0$, they are \emph{local} while for $g = 0$ they are \emph{non-local}.
In other words, the support of $\hat{\mathcal{O}}^{(k)}$ is independent of the system size $L$ while that of ${\hat Q_\mu}$ is extensive in $L$. Consequently, and with the local property of \eqref{eq:ising}, a small deviation from the respective integrable limit results in a \emph{quantum short- or long-range network} spanned by the nonintegrable perturbation among the conserved quantities of the respective integrable limit~\cite{PhysRevResearch.7.023149}. Here we follow the notation developed for classical many-body systems~\cite{srnlrndefinitions2019,merab2022lyapunov}.
For the LRN a local perturbation introduced into the integrable system ($g = 0$) creates couplings between an extensive number of conserved quantities.
Conversely, for the SRN local couplings between conserved quantities are induced by applying local perturbations to the integrable system ($h = 0$).
We emphasize that this classification does not apply to the integrable systems themselves, but rather to the system \emph{infinitesimally close} to the integrable limit.
In the MFIM considered below, the generic integrability-breaking parameter $\lambda$ corresponds to $h$ in the SRN regime and to $g$ in the LRN regime.



%

More precisely, both the SRN and LRN describe the same nonintegrable MFIM, yet they belong to opposite parameter regimes and are well separated by the maximally chaotic regime, which has been characterized by different metrics~\cite{microcanonicalentanglement2024, hamiltonianlearning2026}.
For Eq.~\eqref{eq:ising} with $J=1$, the maximally chaotic regime corresponds to $h \sim g \sim 1$.
Thermalization in a maximally chaotic regime is described by the eigenstate thermalization hypothesis (ETH)~\cite{sred1994chaos,deutsch2018eigenstate}.
In the vicinity of the integrable limit, however, ETH universality is no longer robust~\cite{Mori_2018}.
Several promising directions exist in order to overcome the limitations of ETH, such as full eigenstate thermalization~\cite{foini2019eigenstate1,foini2019eigenstate2}, where a universal theory is developed based on higher-order correlators.
In this work, we focus on a different universality classification, namely that of the \emph{network classes} described above.
In the following sections, we outline the procedures used to probe thermalization and present the results for the SRN and LRN network classes.

We investigate the entanglement relaxation dynamics near both integrable limits of the mixed-field Ising model $\hat{H}(g,h)$~\eqref{eq:ising},
with particular emphasis on the melting of integrable-state plateaus and the possible emergence of intermediate prethermal plateaus.
Fixing $h$ with $g\ll1$ and fixing $g$ with $h\ll1$ correspond to the LRN and SRN regimes, respectively.




\subsection{Entanglement entropy and Melting Time}
\label{sec:entanglement}

To classify thermalization in the MFIM~\eqref{eq:ising} we use the bipartite \emph{entanglement entropy} of an initially random state deduced from the von Neumann entropy of the corresponding reduced density matrix. 
The von Neumann entropy of a quantum state described by a density matrix $\hat{\rho}$ is defined as
\begin{gather}
    S_{\mathrm{vn}} = -\text{Tr}(\hat{\rho}\ln\hat{\rho}).
    \label{eq:vnentropy}
\end{gather}
For a pure state of the full system, $S_{\mathrm{vn}}(\hat{\rho}) = 0$.
Evaluating the von Neumann entropy on the reduced density matrix of one of the subsystems requires fixing a preferred tensor product structure with respect to which bipartition is performed.
In our case, it is the one-dimensional lattice with spins assigned to each site.
The resulting entropy of the reduced state quantifies the quantum correlations between the two complementary parts of the chain.

For an equal bipartition of the chain, this quantity is called the half-chain entanglement entropy, a standard observable in quantum many-body physics~\cite{Amico2008, Calabrese2004}.
Beyond its dynamical growth, which distinguishes thermalization from localization~\cite{DAlessio2016, Abanin2019}, the half-cut entanglement entropy of the eigenstates themselves provides a diagnostic tool to distinguish maximally chaotic systems from integrable ones~\cite{Vidmar2017_Chaotic, Vidmar2017_Integrable}.
While highly excited eigenstates of chaotic Hamiltonians exhibit near-maximal volume-law entanglement consistent with ETH, the macroscopic, conserved quantities in integrable models systematically suppress this volume-law coefficient~\cite{bianchi2022volume}.

We focus on the dynamical evolution of the entanglement entropy.
The density matrix $\hat\rho$ evolving under a Hamiltonian $\hat{H}$, given by $\hat\rho (t)=e^{-i t \hat{H}}\hat\rho e^{i t \hat{H}}$, becomes increasingly entangled across the corresponding bipartition as $t$ increases.
As the system evolves, the entanglement entropy can approach a long-time saturation value.
In strongly chaotic regimes this value is commonly compared with the Page value, whereas near integrable limits additional intermediate or lower saturation plateaus might occur.

Accordingly, the entropy dynamics $S(t)$ is controlled by the Hamiltonian energy differences $\omega_{mn}=E_m-E_n$ and combinations thereof.
To quantify the departure from integrable dynamics, we define, inspired by~\cite{confinedchaos2025}, a characteristic timescale, the melting time $T_M$.
Let $S_{\mathrm{int}}$ denote the long-time entanglement entropy of the corresponding integrable system, evaluated using the same ensemble and averaging procedure.

We choose an entropy threshold $S_M>S_{\mathrm{int}}$ and define $T_M$ as the first time the entropy reaches $S_M$.
When the dynamics exhibit a second, distinct intermediate plateau, we introduce an additional threshold $S_{M_2}>S_M$ and define the second melting time $T_{M_2}$ through its first crossing.
If the second threshold lies above an intermediate prethermal plateau, $T_{M_2}$ characterizes its subsequent melting. The interpretation of these timescales depends on the structure of the entropy dynamics.
In particular, the departure from the integrable plateau does not, by itself, establish the presence of a prethermal plateau.
Moreover, when the growth is nonuniform, the extracted melting times might depend on the thresholds chosen.

We then analyze $T_M$ as a function of the integrability-breaking parameter $\lambda$.
A power-law dependence $T_M\propto\lambda^{-\alpha}$ is commonly expected~\cite{Langen_2016, Mori_2018}, and in some cases the exponent $\alpha=2$ can be derived or argued starting from Fermi's golden rule~\cite{prethermalizationrigol2021}.
However, deviations from this simple algebraic dependence might occur.
Deviations from the exponent \(2\) have been observed in both classical~\cite{confinedchaos2025} and quantum~\cite{prethermalbukov2021} systems.
Furthermore, under certain conditions a spectrum of exponents can emerge in the context of weak integrability-breaking perturbations~\cite{prethermalstaircase2025, weakintegrability2023}.

\section{Results}
\label{sec:results}

We consider a chain of $L$ spins with the equal bipartition $\mathscr{H} = \mathscr{H}_{A}\otimes\mathscr{H}_{B}$, where $L_A$ and $L_B$ denote the numbers of spins in subsystems $A$ and $B$, respectively.
To ensure equal dimensions of the subsystems~\footnote{The results do not depend qualitatively on the choice of bipartition in the lattice basis}, we consider only even $L$.
An initial state is evolved unitarily as $\ket{\psi(t)} = e^{-i\hat{H}(g,h)t}\ket{\psi(0)}$.
The reduced density matrix of subsystem $A$ is
\begin{gather}
    \hat{\rho}_{A}(t) = \mathrm{Tr}_{B}\left\{\hat{U}(t)\hat{\rho}\hat{U}^{\dagger}(t)\right\} \,, \quad \hat{\rho} = \ketbra{\psi(0)}{\psi(0)} \,,
    \label{eq:reducedrho}
\end{gather}
where $\ket{\psi(0)}=\ket{\theta_1,\phi_1}\otimes\ket{\theta_2,\phi_2}\otimes\cdots\otimes\ket{\theta_L,\phi_L}$ is a random product state (RPS) in which each spin $\ket{\theta_i,\phi_i}$ points in a random direction on the Bloch sphere defined by the angles $(\theta_i,\phi_i)$.
The von Neumann entropy is given by
\begin{gather}
    S(t) = -\sum_{i=1}^{2^{L_{A}}}\lambda_{i}(t)\ln[\lambda_{i}(t)] \, ,
    \label{eq:entanglement}
\end{gather}
where $\lambda_i(t)$ are the eigenvalues of $\hat{\rho}_{A}(t)$.
An ensemble of initial RPSs was used and a running mean was applied because the entanglement entropy of individual random states is highly oscillatory~\cite{temporalfluc2023,PhysRevResearch.1.033067}.
In the energy eigenbasis, the initial state is $\ket{\psi(0)}=\sum_i c_i\ket{E_i}$, and the reduced density matrix takes the form

\begin{gather}
    \hat{\rho}_A(t) = \sum_{i,j}c_{i}c_{j}^{*}e^{-it(E_{i}-E_{j})}\mathrm{Tr}_{B}\left\{\ketbra{E_{i}}{E_{j}}\right\} \,.
    \label{eq:rhoexpansion}
\end{gather}

Its time evolution is thus determined by the energy differences $E_i - E_j$.
Which differences contribute depends on the non-zero coefficients $c_i = \braket{E_i}{\psi(0)}$, and the weight of those contributions also depends on the entanglement structure of the eigenstates.
Consequently, the weight of the contributions of the energy differences to the reduced density matrix depends on the choice of initial states.
The use of RPS ensures that nearly all energy differences appear in the sum and that there is no initial entanglement across any bipartition of the chain.
We have also compared RPS with product states of the form $\ket{\psi(0)}=\ket{\psi_A}\otimes\ket{\psi_B}$, where $\ket{\psi_A}$ and $\ket{\psi_B}$ are independently sampled from the Haar measure on $\mathscr{H}_{A}$ and $\mathscr{H}_{B}$, respectively;
the results are qualitatively similar.

We analyze the structure of the energy levels at the integrable limits and compute the entanglement entropy for an ensemble of \(50\) RPS.
Using exact diagonalization, we evaluate the reduced density matrix and the corresponding time-dependent half-chain entanglement entropy through Eqs.~\eqref{eq:reducedrho} and~\eqref{eq:entanglement}. To reduce the oscillatory behavior inherent to the entropy for individual RPS~\cite{temporalfluc2023}, we perform an ensemble average followed by a running mean over a window of $200$ points, with the data sampled at 1000 logarithmically spaced points per decade in time.
All entanglement-dynamics figures report only this processed quantity.

The melting times are extracted from threshold crossings of the running-mean ensemble-averaged entanglement entropy. For a given integrability-breaking parameter $\lambda$, we determine the scaling of these timescales through power-law fits $T_{M_j}\propto\lambda^{-\alpha_j}$, where $j$ distinguishes the successive thresholds when more than one is considered.
The corresponding entropy thresholds are specified separately for each numerical analysis.

In the LRN case, three values $h = 0.1, 0.3, 1.0$ were chosen for system size $L=8$, spanning a range from nearly degenerate spectra to spectra with partially lifted degeneracy.
In the SRN regime, we focus on the critical value $g=2$ and use $g=1$ as a representative value to illustrate the behavior observed at the other values of $g$ examined.
All SRN results were checked for $L = 6, 8, 10, 12, 14$, with no systematic size dependence within this range.

\subsection{SRN Regime}
\label{sec:srn}

\begin{figure}[htbp]
    \centering
    \includegraphics[width=\columnwidth]{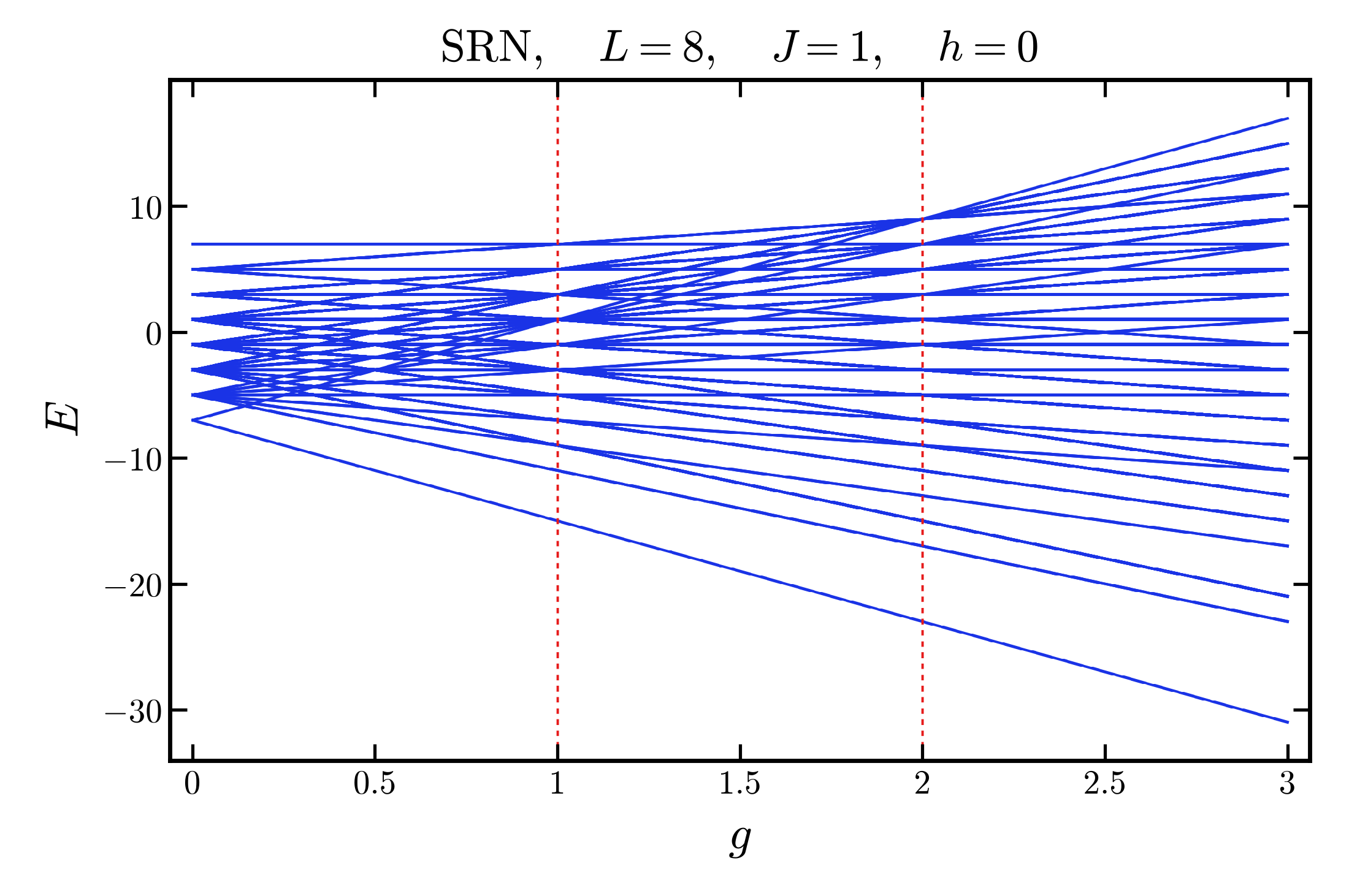}
    \caption{
        Parametric evolution of the energy levels for the integrable SRN case as a function of $g$.
        The two vertical red dotted lines correspond to the values $g = 1$ and $g = 2$ used in the numerical analysis.
    }
    \label{fig:srnlevels}
\end{figure}

As discussed previously, the SRN limit corresponds to the Hamiltonian~\eqref{eq:ising} with $h \ll 1$.
Figure~\ref{fig:srnlevels} shows the energy levels in the integrable SRN limit as a function of $g$.
This spectrum is the starting point for the prethermalization analysis.

At the integrable limit \(h=0\), the Hamiltonian $\hat{H}$ is diagonal in the computational basis. 

Since we measure only the entanglement entropy generated across the equal bipartition, the microscopic details of the initial state within each subsystem are irrelevant.
The sole source of entanglement is the bond coupling the two spins adjacent to the cut.
From an operator point of view, the Hamiltonian and unitary evolution factorize as
\begin{align*}
    \hat{H}_{h=0} &= \hat{H}_A \otimes \hat{I}_B + \hat{I}_A \otimes \hat{H}_B + \hat{H}_I\,, \\
    \hat{U}(t) &= (e^{-i\hat{H}_At} \otimes e^{-i\hat{H}_Bt})\,e^{-i\hat{H}_I t}\,.
\end{align*}
The local unitaries $e^{-i\hat{H}_At}$ and $e^{-i\hat{H}_Bt}$ cannot generate entanglement across the cut,
whereas this is possible for the interaction term $\hat{H}_I=-J\hat{\sigma}_{L/2}^{z}\hat{\sigma}_{L/2+1}^{z}$, which acts on exactly one qubit in each part.
This limits the Schmidt rank of the evolved state to at most two, and therefore, independently of system size, the entanglement entropy at the integrable limit is bounded by $\ln(2)$.

The above analysis breaks down once the perturbation $\lambda\hat{V} = -h\sum_i\hat{\sigma}_{i}^x$ is added, allowing the entanglement entropy to grow beyond $\ln(2)$.
Note that this perturbation breaks all conserved quantities simultaneously. However, the integrability breaking, quantified by the commutator $\left[h\sum_i \sigma^{x},\hat{\mathcal{O}}^{(k)}\right]$ is local.
Therefore, the effect on the dynamics of the operators originating from such nested commutators is local for small $t, h$.
It is precisely this breakdown of exact integrability that is quantified by the melting time $T_M$.
Since the entanglement dynamics always depend on the initial state and the activated energy differences, we present results for two distinct values of $g$ at the integrable limit.

\begin{figure}[htbp]
    \centering
    \includegraphics[width=\columnwidth]{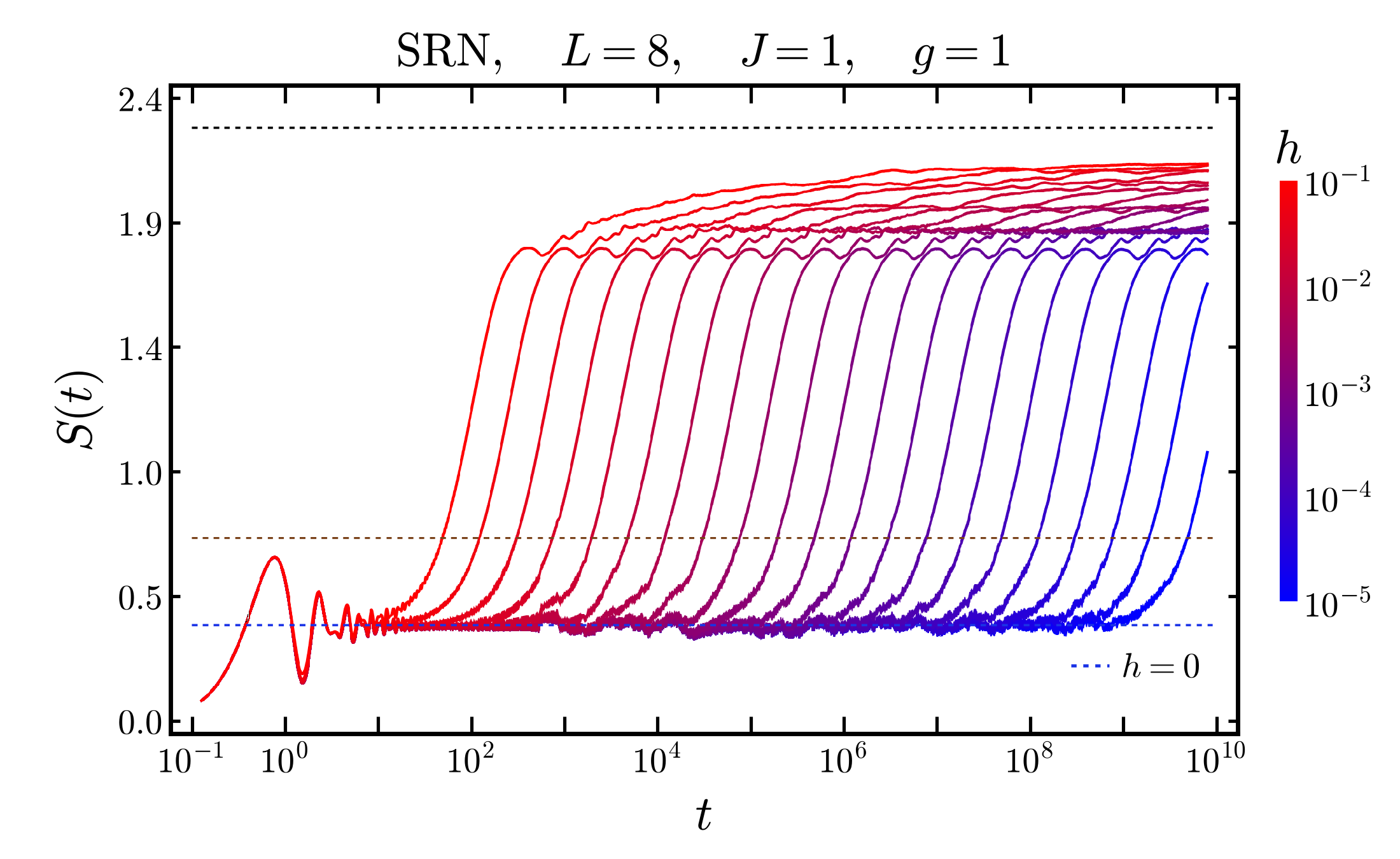}
    \caption{
        SRN analysis of Fig.~\ref{fig:srnlevels} for $g=1$.
        The figure shows the running mean of the entanglement entropy averaged over \(50\) random product states for different values of the integrability-breaking parameter $h$, ranging from $h = 10^{-5}$ to $h = 10^{-1}$, logarithmically spaced with step $\Delta\log_{10} h = 0.2$.
        The melting time $T_M$ is the first time the running-mean entropy reaches $S_M=0.7$, marked by the brown dotted line.
        The blue dotted line indicates the steady value of the running mean of the entanglement entropy at integrability ($h=0$) and the black dotted line marks the Page value.
    }
    \label{fig:srnentropyg1}
\end{figure}

At $g = 1$, a power-law fit of the melting time yields $T_M \propto h^{-2}$.
We obtain the same exponent at the other values of $g$ examined away from $g=2$. We observe no systematic dependence on $L$ for the system sizes studied, $L=6,8,10,12,14$ (not shown).
Figure~\ref{fig:srnentropyg1} shows the evolution of the entanglement entropy for an ensemble of \(50\) random product states.
At the integrable limit, the ensemble-averaged entropy reaches a stationary value bounded by $\ln 2$, consistent with the preceding analytical argument.
This provides the integrable reference plateau.
When the perturbation is introduced, the entropy departs from this reference value.

Because the departure is homogeneous across system sizes and realizations, a threshold value of $S = 0.7$ is used to extract $T_M$. At $h = 0$, the conserved quantities given in~\eqref{eq:SRNcharges} are strictly local. For sufficiently small $h \neq 0$, they can be perturbatively dressed into quasi-local operators $\hat{\tau}^z_i = \hat{\sigma}^z_i + \mathcal{O}(h)$.
These are analogous to the local integrals of motion of many-body localized systems, known as l-bits, whose dephasing leads to a slow, unbounded growth of entanglement~\cite{PhysRevLett.111.127201,PhysRevB.90.174202,srnmbl1,srnmbl2}. Unlike in those disordered systems, however, the operators $\hat{\tau}^z_i$ are only approximately conserved here: they decay at a rate that vanishes as $h \to 0$, so the dynamics remains constrained by them for increasingly long times~\cite{Mori_2018,Prethermalizationchaotic2019,prethermalizationrigol2021}.


Since the exact-diagonalization dynamics is limited by the eigenvalue precision, $\delta E \sim \epsilon_{\mathrm{machine}} \lVert \hat{H} \rVert$, where $\epsilon_{\mathrm{machine}} \approx 2.2\times10^{-16}$ is the double-precision machine epsilon, the phases $e^{-iE_n t}$ remain reliable only up to $t \lesssim 10^{13}$--$10^{14}$.
Consequently, the asymptotic saturation is reached within the accessible time window only for the largest perturbations.

For this case, the choice of the threshold $S_M=0.7$ is not highly restrictive, since the initial entropy growth is approximately
uniform across the perturbation strengths considered, up to entropies around $S=1.7$ for $L=8$.
At later times, the growth slows down before reaching its final saturation.
The resulting scaling $T_M\propto h^{-2}$ therefore characterizes the melting of the integrable-state plateau.
Although the subsequent growth is slower, Fig.~\ref{fig:srnentropyg1} does not establish a distinct, long-lived prethermal plateau.

\begin{figure}[htbp]
    \centering
    \includegraphics[width=\columnwidth]{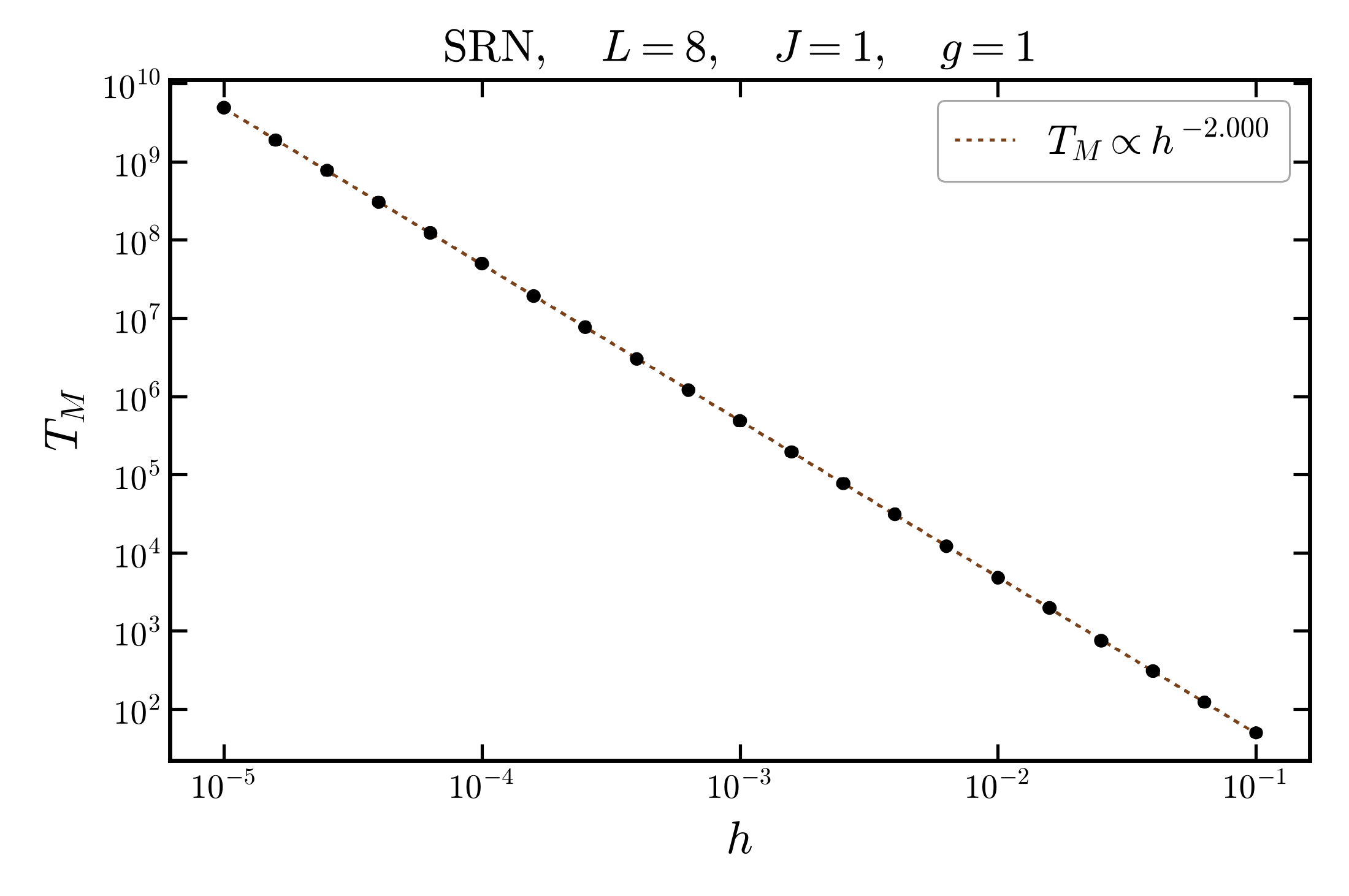}
    \caption{
        Melting time $T_M$ extracted from Fig.~\ref{fig:srnentropyg1}. The data follow $T_M\propto h^{-2}$. We observe no systematic size dependence for $L=6,8,10,12,14$ (not shown). The same scaling holds at the other values of $g$ examined away from $g=2$.}
    \label{fig:srnpowerg1}
\end{figure}

\begin{figure}[htbp]
    \centering
    \includegraphics[width=\columnwidth]{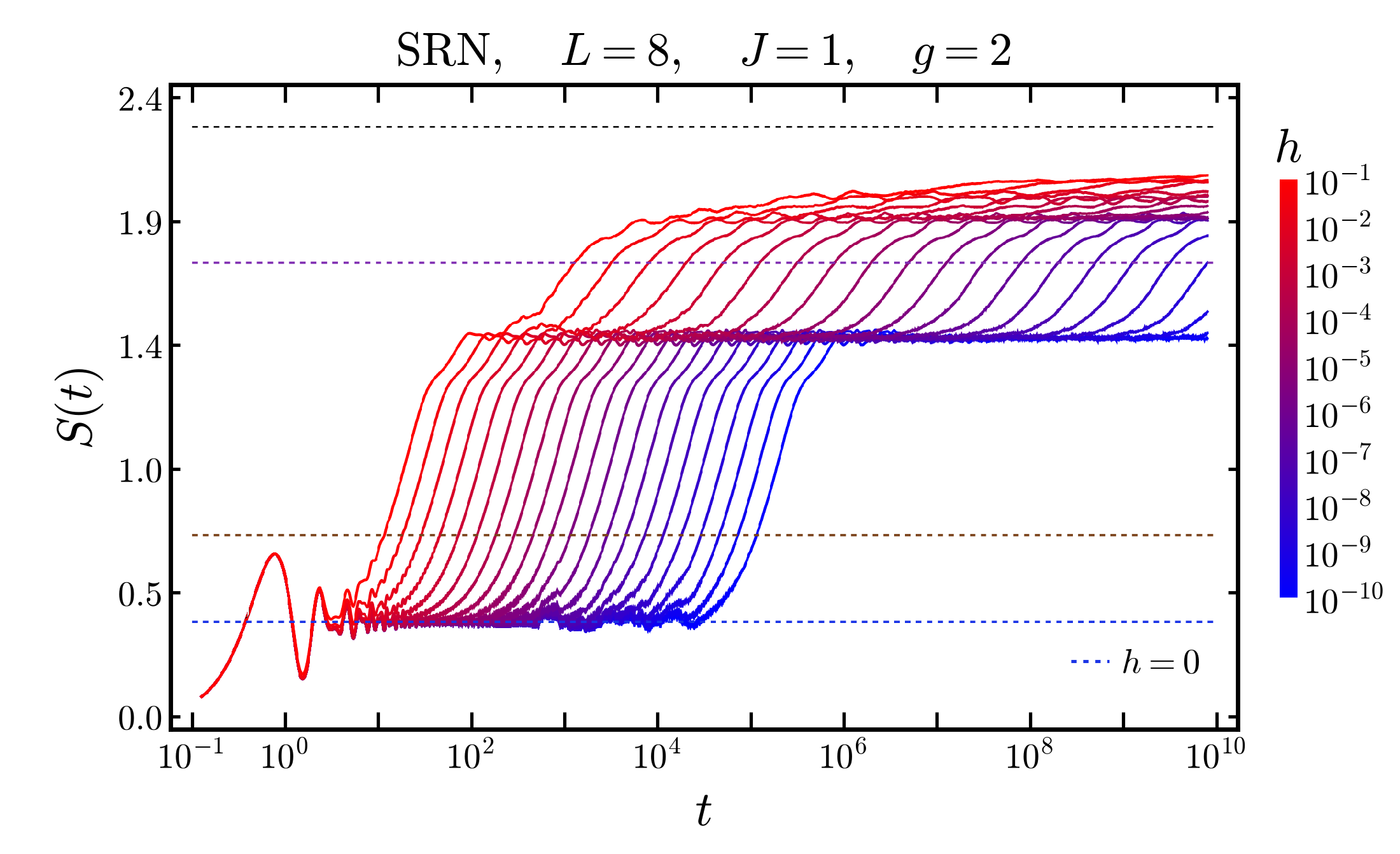}
    \caption{
        SRN analysis at the critical value $g=2$.
        Running-mean entanglement entropy averaged over \(50\) random product states for perturbations $h=10^{-10}$ to $10^{-1}$, with
        $\Delta\log_{10}h=0.2$.
        The brown dotted line at $S_M=0.7$ defines the first melting time $T_M$, measuring the departure from the integrable-state plateau.
        The purple dotted line at $S_{M_2}=1.75$ defines $T_{M_2}$, measuring the subsequent melting of the intermediate prethermal plateau.
        The blue dotted line indicates the integrable-limit reference entropy ($h=0$), and the black dotted line denotes the Page value.
    }
    \label{fig:srnentropyg2}
\end{figure}

\begin{figure}[h!]
    \centering
    \includegraphics[width=\columnwidth]{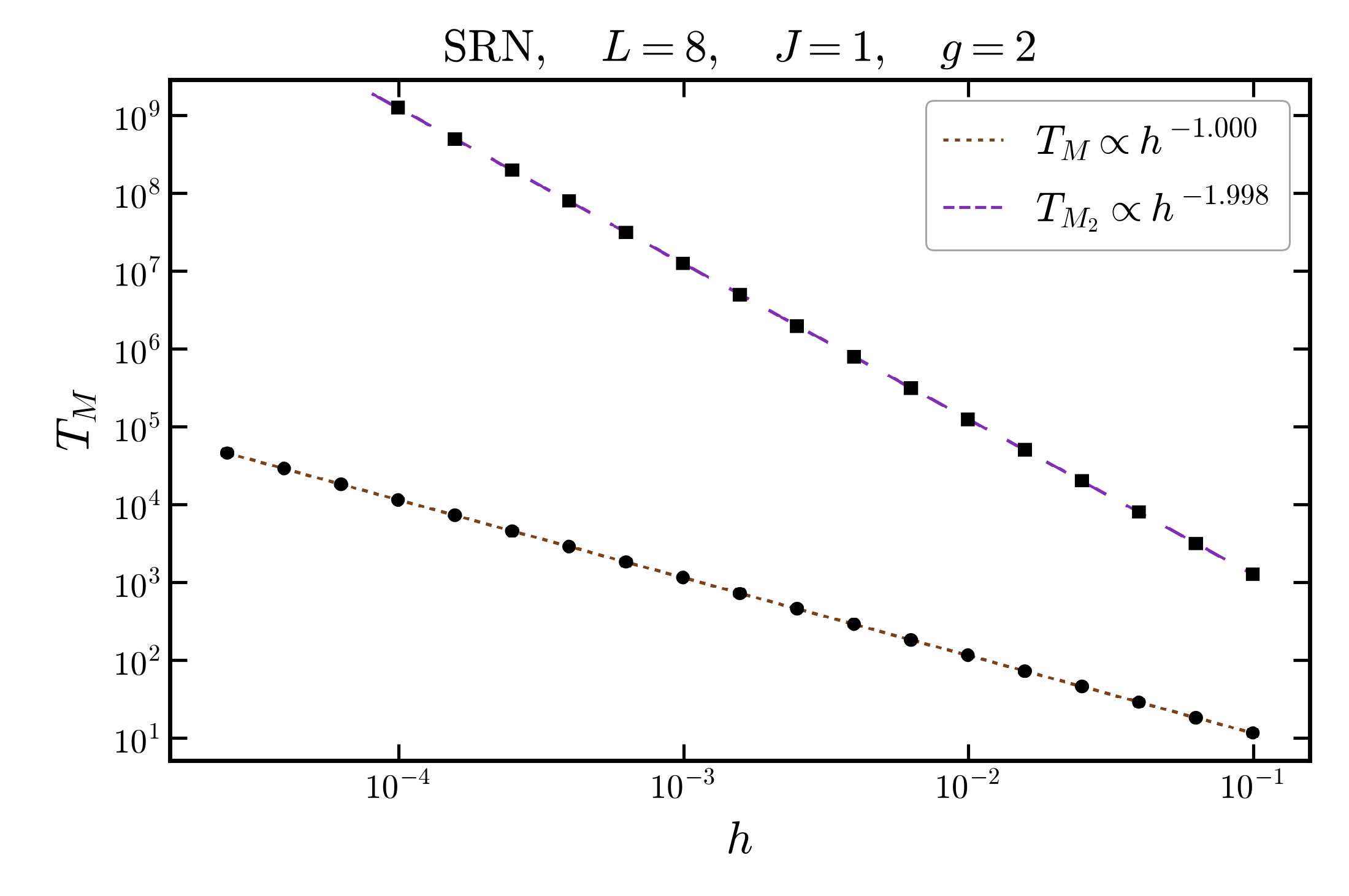}
    \caption{
        Two melting times extracted from Fig.~\ref{fig:srnentropyg2} for SRN at $g=2$.
        The first timescale follows $T_M\propto h^{-1.000}$ and characterizes the departure from the integrable-state plateau.
        The second follows $T_{M_2}\propto h^{-1.998}$ and characterizes the subsequent melting of the intermediate prethermal plateau.
        Black circles and squares indicate the numerically extracted times, while the brown dotted and purple dashed lines indicate the corresponding power-law fits.
    }
    \label{fig:srnpowerg2}
\end{figure}

At the critical value $g=2$~\cite{critical1fibonacci,critical2}, the entanglement dynamics exhibit a qualitatively different relaxation structure, as shown in Figs.~\ref{fig:srnentropyg2} and~\ref{fig:srnpowerg2}.
The initial departure from the integrable-state plateau is characterized by the threshold $S_M=0.7$, yielding the scaling $T_M\propto h^{-1.0}$.
However, unlike the case $g=1$, the entropy does not evolve directly toward its final saturation value.
Instead, it reaches a second, clearly distinguishable intermediate plateau.
This intermediate plateau constitutes the prethermal regime identified in the present analysis.
To characterize its subsequent melting, we introduce a second threshold, $S_{M_2}=1.75$, above the intermediate plateau.
The resulting melting time follows $T_{M_2}\propto h^{-1.998}$, as shown in Fig.~\ref{fig:srnpowerg2}.

We therefore distinguish two successive processes:
the initial melting of the integrable-state plateau, governed by an exponent close to $1$, and the subsequent melting of the intermediate prethermal plateau, governed by an exponent close to $2$.
The separation of these processes is particularly evident as the perturbation strength decreases.
The resulting multistage dynamics is reminiscent of staircase prethermalization in lattice gauge theories with weakly
broken local symmetries ~\cite{halimeh2020staircase,prethermalstaircase2025}.
Related prethermal regimes have also been investigated in periodically kicked spin chains~\cite{prethermalbukov2021} and in systems with weak integrability-breaking perturbations ~\cite{weakintegrability2023}.
In the present case, the dynamics arise from a time-independent Hamiltonian, without Floquet driving or Trotterization.
The microscopic mechanism responsible for the intermediate prethermal plateau remains to be established.

\subsection{LRN regime}
\label{sec:lrn}

\begin{figure}[htbp]
    \centering
    \includegraphics[width=\columnwidth]{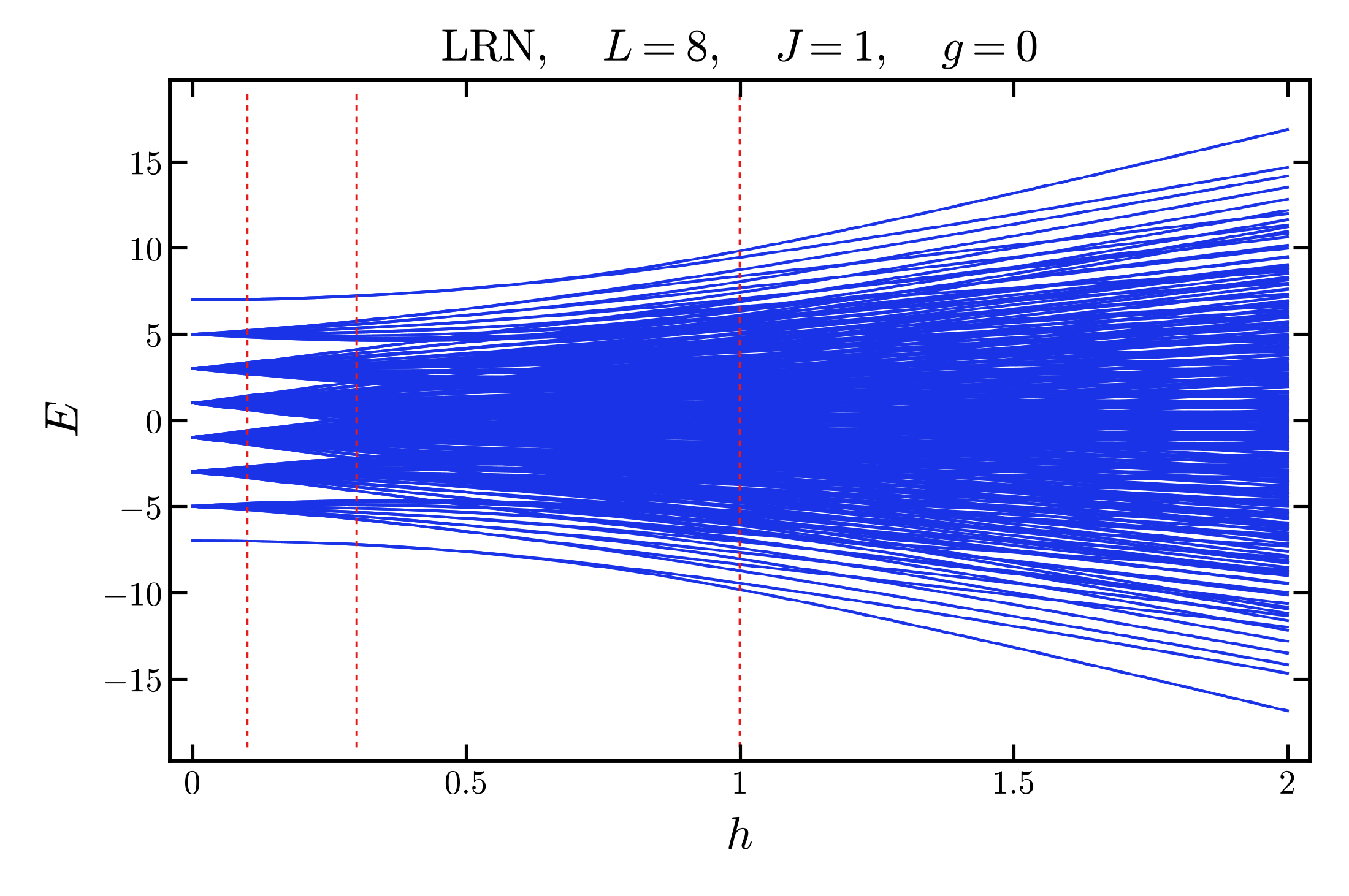}
    \caption{
        Parametric evolution of the energy levels at the integrable LRN limit ($g=0$) as a function of the transverse field $h$ for $L=8$.
        The three vertical red dotted lines indicate $h=0.1$, $0.3$, and $1.0$, corresponding to the values used in the subsequent dynamical analysis.
    }
    \label{fig:lrnlevels}
\end{figure}

Figure~\ref{fig:lrnlevels} shows the parametric evolution of the energy levels in the integrable limit $g=0$ as a function of $h$.
The three values $h = 0.1, 0.3, 1.0$ used in the numerical analysis are indicated by the red dotted lines.

{The longitudinal perturbation} $\lambda\hat{V}=-g\sum_i\hat\sigma_i^z$ {acts in the spin basis, in which the conserved quantities in Eq.~\eqref{eq:LRNcharges} have extended support,} {implying that for $h\ne0$, each term $-\hat\sigma_i^z$ acts nontrivially on all of them.} Indeed, for this case integrability can be broken even by a single defect~\cite{leaspeck2020}. Consequently, determining the maximum entanglement entropy reachable by the perturbed TFIM starting from an RPS is non-trivial. We have verified that, for a fixed system size, at the integrable limit $g=0$ the entanglement entropy grows initially and then saturates well below the Page value, consistent with the proven bound for free fermionic Hamiltonians~\cite{bianchi2022volume,bianchi2021page,iannotti2025entanglement}. This saturation defines the first entanglement plateau;
in the LRN regime, the melting time $T_M$ measures the departure from this plateau upon adding the perturbation.


The final saturation value after integrability is broken cannot be unambiguously determined from the present data.
For the largest perturbations, the entropy saturates within the simulated time window, but below the Page value, consistent with the energy-conservation constraint discussed in Refs.~\cite{microcanonicalentanglement2024,PhysRevD.100.105010}.
For smaller perturbations, the entropy instead settles on lower, $g$-dependent plateaus (see Appendix), and our data do not distinguish whether these are asymptotic values or intermediate plateaus that would continue to grow beyond the simulated window.
We therefore do not assign a characteristic final saturation value in the LRN regime.


\subsubsection{$h = 1$}

\begin{figure}[htbp]
    \centering
    \includegraphics[width=\columnwidth]{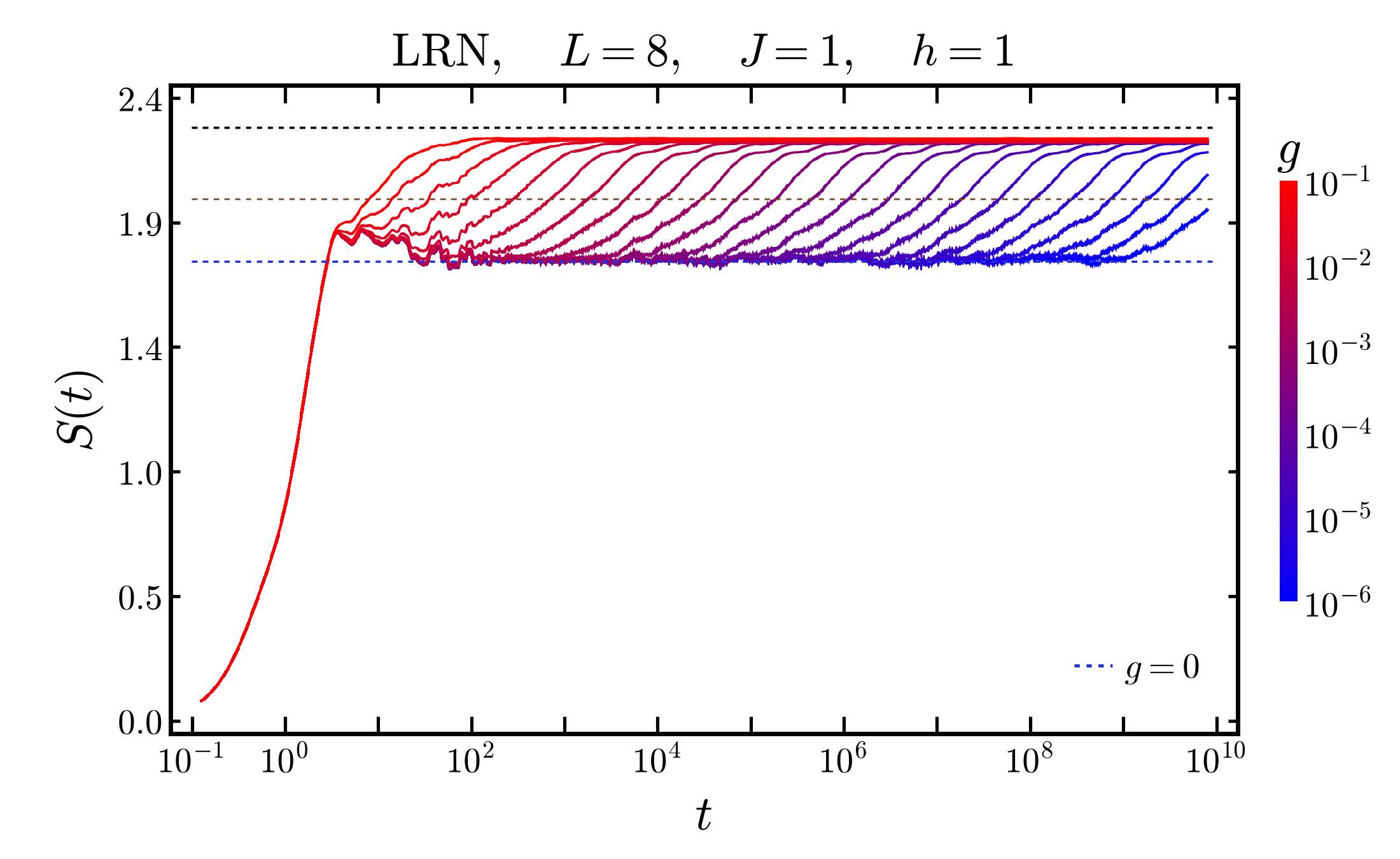}
    \caption{
        LRN analysis of Fig.~\ref{fig:lrnlevels} for $h=1$.
        The figure shows the running mean of the entanglement entropy averaged over \(50\) random product states for different values of the integrability-breaking parameter $g$, ranging from $g = 10^{-6}$ to $g = 10^{-1}$, logarithmically spaced with step $\Delta\log_{10} g = 0.2$.
        The melting time $T_M$ is the first time the running-mean entropy reaches $S_M=2.0$, marked by the brown dotted line.
The blue dotted line marks the integrable reference entropy ($g=0$), and the black dotted line marks the Page value.
    }
    \label{fig:lrnentropyh1}
\end{figure}

\begin{figure}[htbp]
    \centering
    \includegraphics[width=\columnwidth]{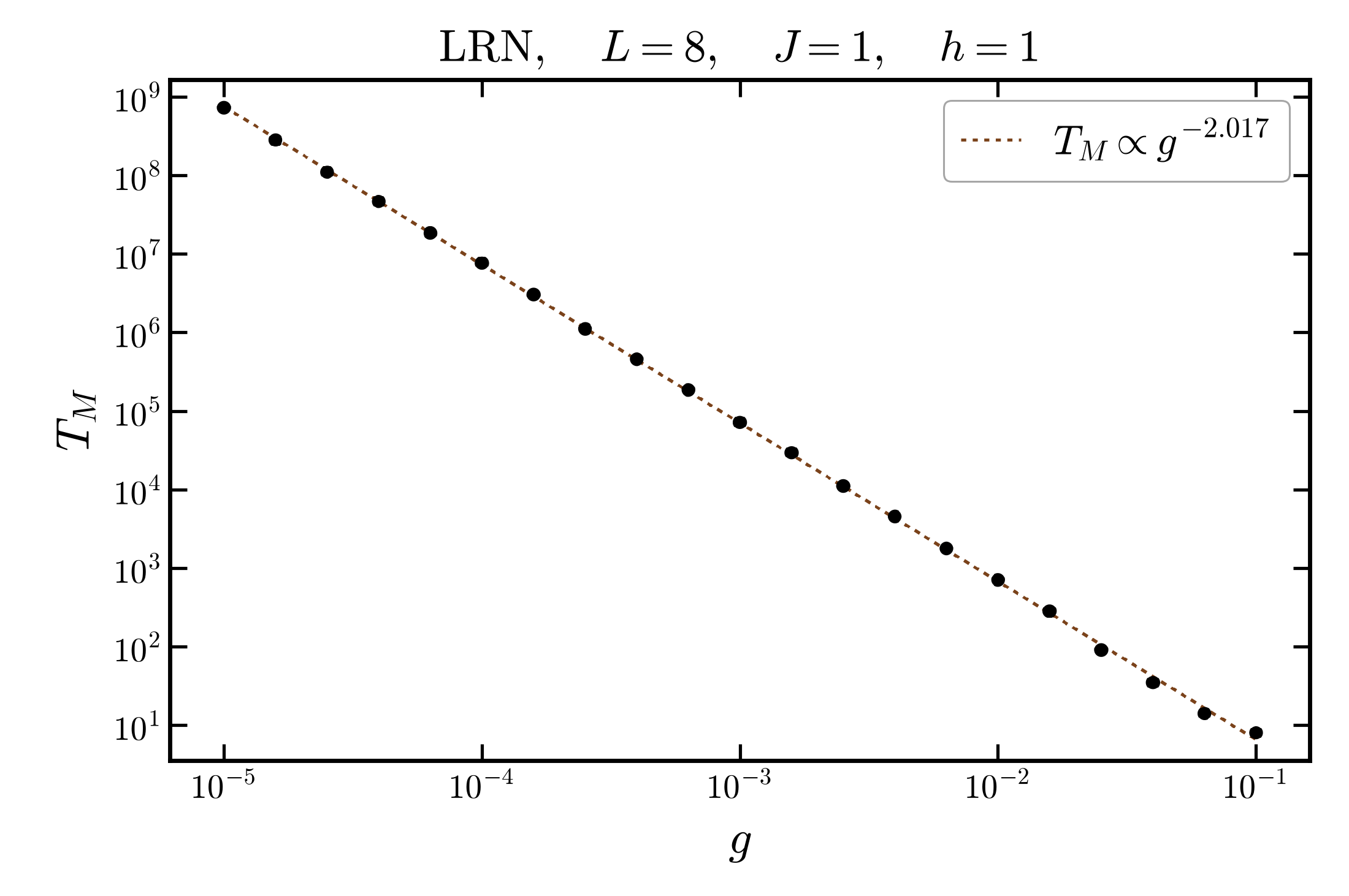}
    \caption{
        Melting time $T_M$ extracted from Fig.~\ref{fig:lrnentropyh1}.
        The brown dotted line is a power-law fit with the exponent indicated in the inset.
    }
    \label{fig:lrnpowerh1}
\end{figure}

At $h = 1$, nearly all unperturbed energy levels (\(256\) in total from both reflection sectors) form a single bulk. 
Figures~\ref{fig:lrnentropyh1} and~\ref{fig:lrnpowerh1} show the entanglement entropy evolution and the extracted melting time, respectively.
The power-law fit yields $T_M \propto g^{-2.017}$.

The key qualitative difference between the LRN and SRN regimes lies in how the shape of the temporal entropy growth changes with the perturbation strength. In the SRN regime away from $g=2$ (Fig.~\ref{fig:srnentropyg1}), the curves for different $h$ are, to a good approximation, rigidly shifted along the logarithmic time axis. We obtain the same melting-time exponent over the thresholds tested in this range.
In the LRN regime this is not the case: as $g$ decreases, the growth from the integrable plateau spreads over an increasing time interval and, for some parameters, develops intermediate shoulders (most clearly for $L=10$ and $12$; see Appendix~\ref{app:lrnscaling}).
Because the curves do not collapse under a shift in $\log t$, different thresholds probe different stages of the growth, and the extracted exponent becomes threshold dependent, as discussed at the end of this section.
Increasing the number of initial states in the ensemble or the number of sampling points in the time evolution does not remove this effect, indicating that it is not a statistical artifact.


\subsubsection{$h = 0.1$}

\begin{figure}[htbp]
    \centering
    \includegraphics[width=\columnwidth]{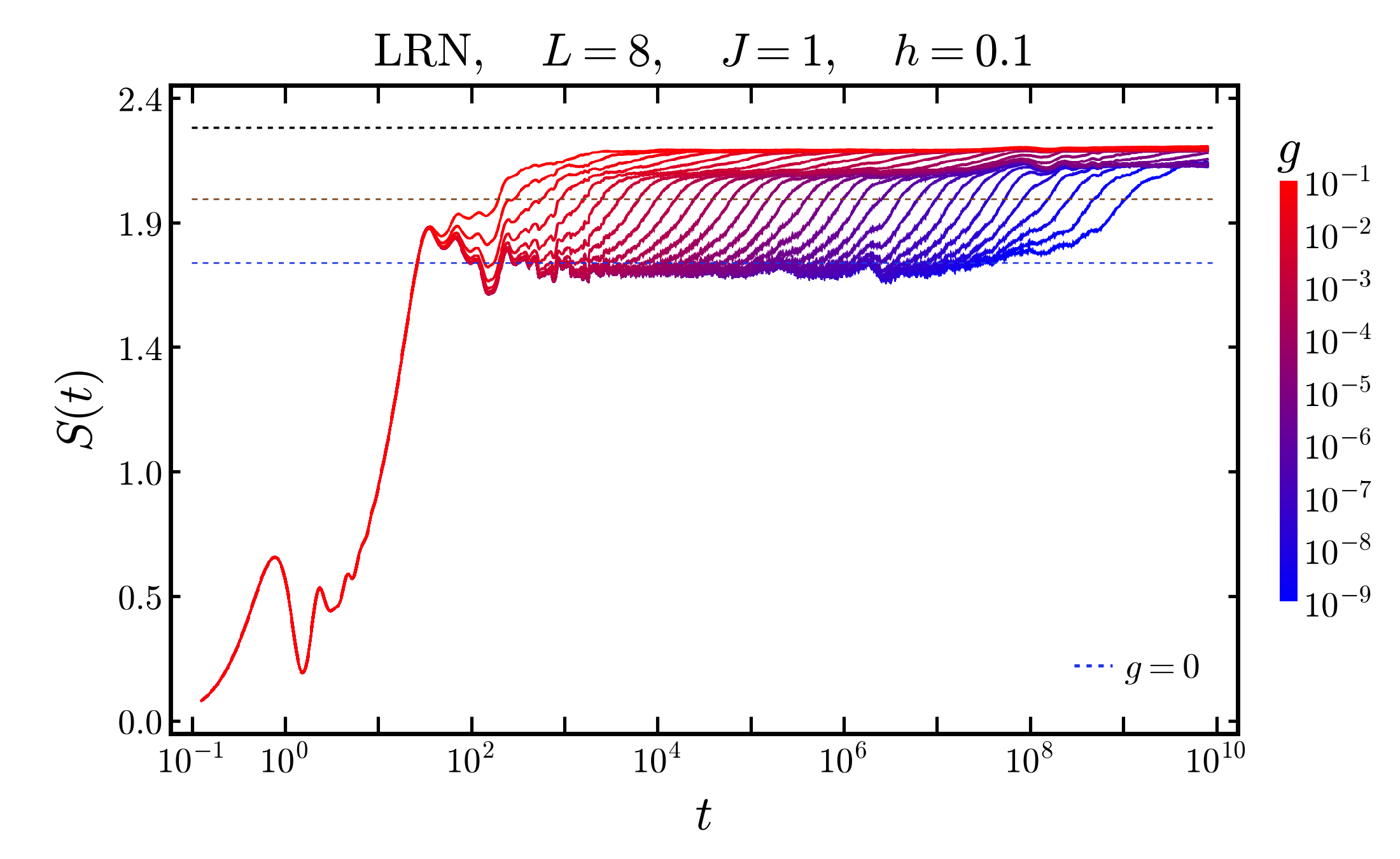}
    \caption{LRN analysis of Fig.~\ref{fig:lrnlevels} for $h = 0.1$.
        The figure shows the running mean of the entanglement entropy averaged over \(50\) random product states for different values of the integrability-breaking parameter $g$, ranging from $g = 10^{-9}$ to $g = 10^{-1}$, logarithmically spaced with step $\Delta\log_{10} g = 0.2$.
        The melting time $T_M$ is the first time the running-mean entropy reaches $S_M=2.0$, marked by the brown dotted line.
The blue dotted line marks the integrable reference entropy ($g=0$),
and the black dotted line marks the Page value. The early-time entropy dynamics resemble those in the SRN regime shown in Fig.~\ref{fig:srnentropyg1}. The power-law fit uses perturbation strengths in the interval
$g\in[10^{-8.2},10^{-2.2}]$.
    }
    \label{fig:lrnentropyh01}
\end{figure}

At $h=0.1$, the spectrum forms tight clusters, as shown in Fig.~\ref{fig:lrnlevels}. The levels within each cluster approach degeneracy as $h\to0$. The power-law fit yields $T_M \propto g^{-1.005}$. The early-time dynamics of the entanglement entropy resemble the SRN case (cf.\ Fig.~\ref{fig:srnentropyg1}), a similarity that strengthens at smaller values of $h$.
An additional region of slow entropy growth is also visible in Fig.~\ref{fig:lrnentropyh01}.
However, its identification as a distinct prethermal plateau and the characterization of its subsequent lifetime remain unresolved.
We therefore restrict the present analysis to the melting time extracted from the first threshold.

\begin{figure}[htbp]
    \centering
    \includegraphics[width=\columnwidth]{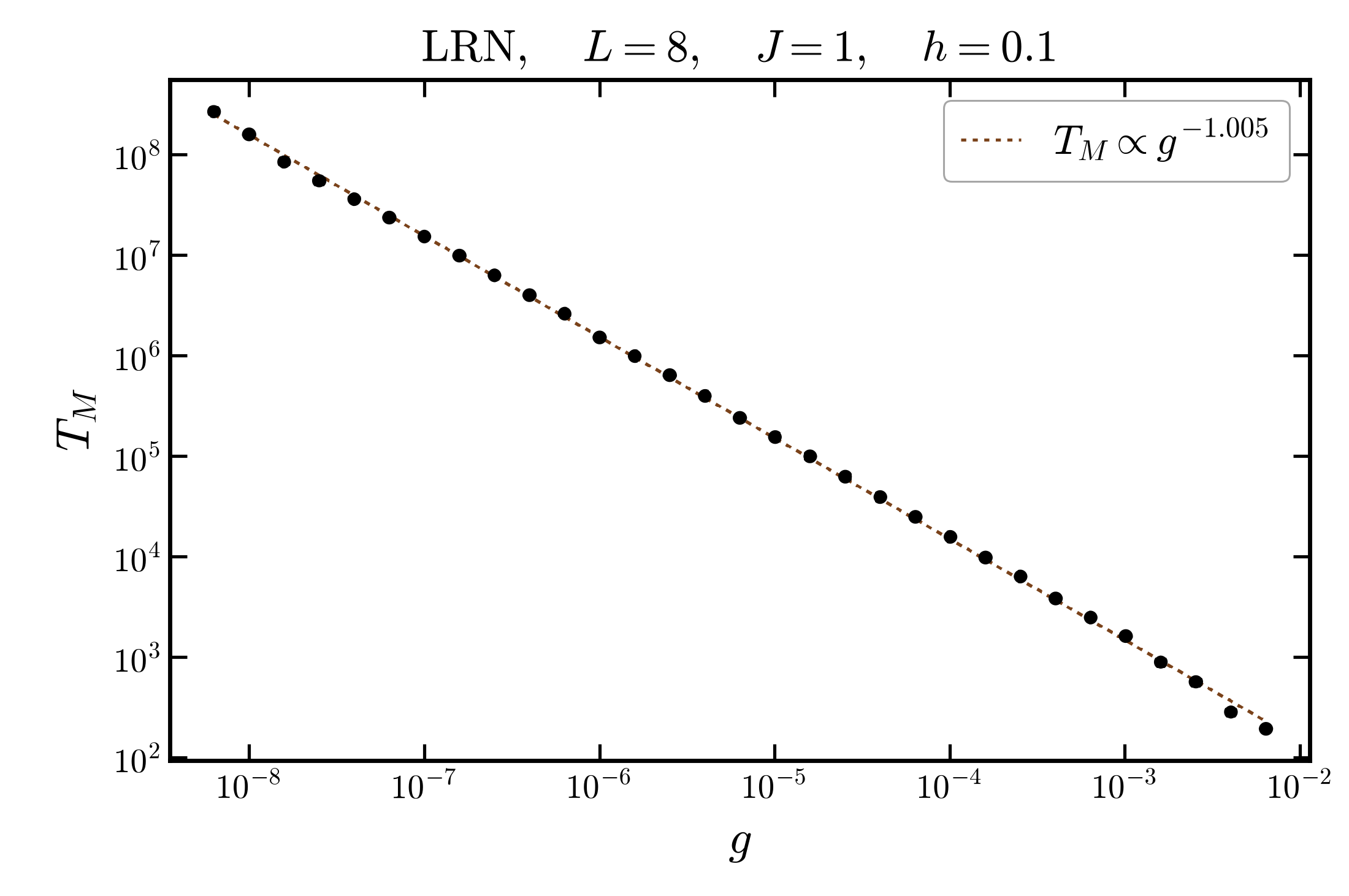}
    \caption{
        Melting time $T_M$ extracted from Fig.~\ref{fig:lrnentropyh01}.
        The brown dotted line is a power-law fit with the exponent indicated in the inset.
    }
    \label{fig:lrnpowerh01}
\end{figure}

\subsubsection{Crossover region, $h = 0.3$}

At $h=0.3$, the spectrum lies between the clustered and overlapping
regimes shown in Fig.~\ref{fig:lrnlevels}.
The melting time follows $T_M\propto g^{-1.985}$ over
$g\in[10^{-6.8},10^{-4.8}]$ and $T_M\propto g^{-1.055}$ over
$g\in[10^{-4.8},10^{-2}]$.

\begin{figure}[h!]
    \centering
    \includegraphics[width=\columnwidth]{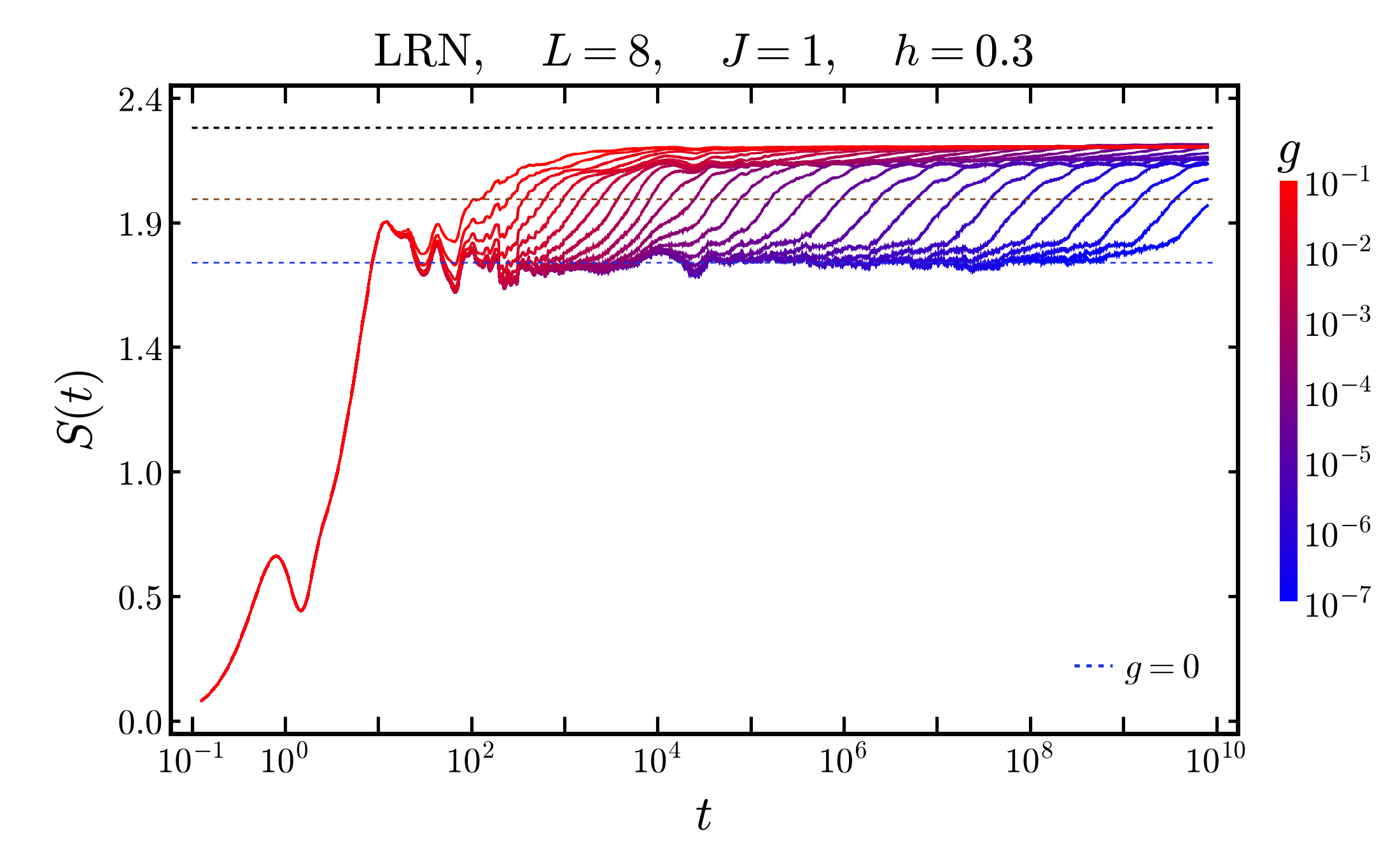}
    \caption{LRN analysis of Fig.~\ref{fig:lrnlevels} for $h = 0.3$.
        The figure shows the running mean of the entanglement entropy averaged over \(50\) random product states for different values of the integrability-breaking parameter $g$, with a logarithmic color scale spanning $10^{-7}$ to $10^{-1}$.
        The melting time $T_M$ is the first time the running-mean entropy reaches
$S_M=2.0$, marked by the brown dotted line.
The blue dotted line marks the integrable reference entropy ($g=0$),
and the black dotted line marks the Page value.
        For this case, the two fits use $g\in[10^{-6.8},10^{-4.8}]$ and $g\in[10^{-4.8},10^{-2}]$, respectively.
    }
    \label{fig:lrnentropyh03}
\end{figure}

\begin{figure}[t]
    \centering
    \includegraphics[width=\columnwidth]{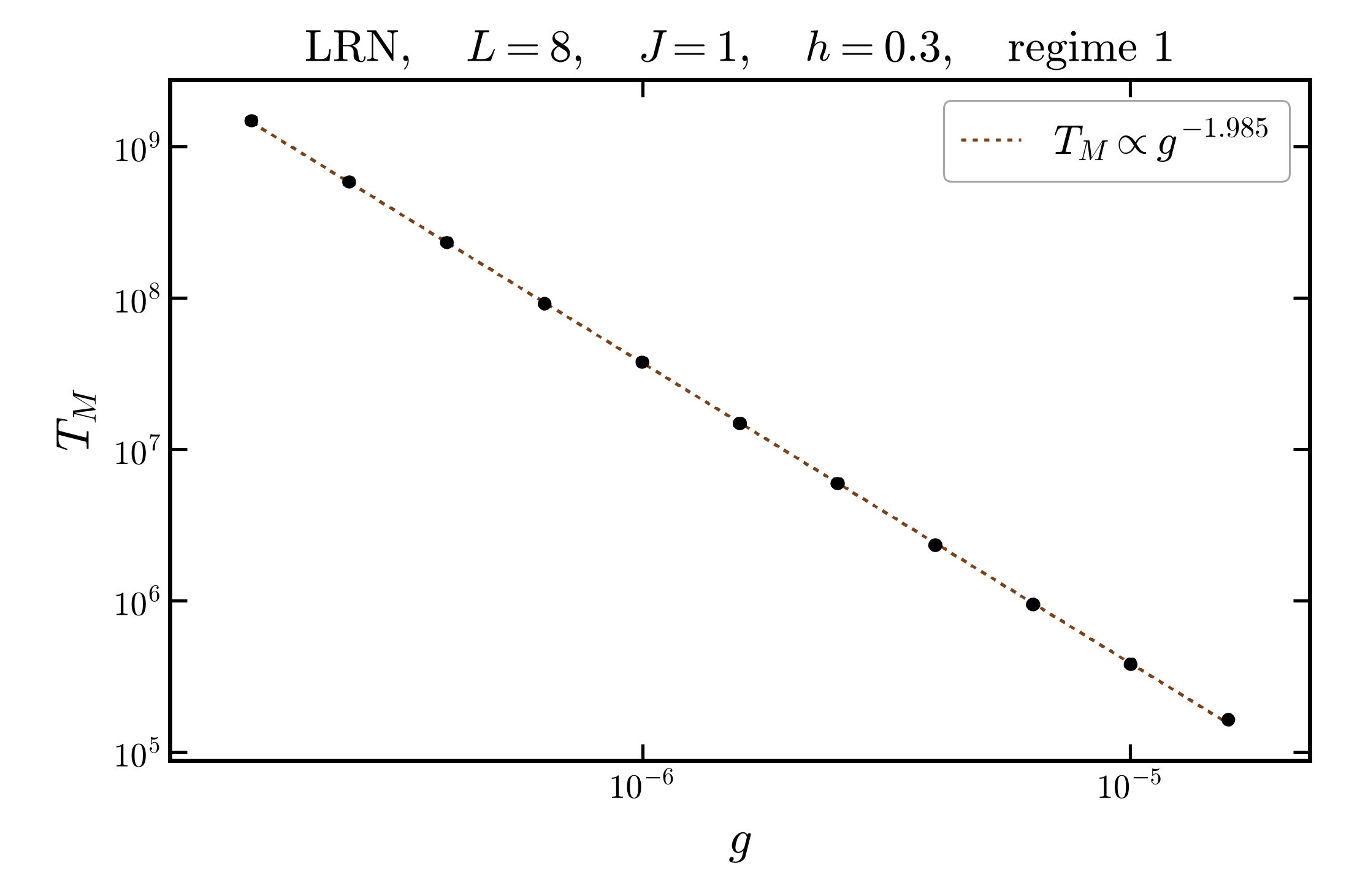}
    \caption{
        First regime: melting time $T_M$ extracted from Fig.~\ref{fig:lrnentropyh03} for $g\in[10^{-6.8},10^{-4.8}]$.}
    \label{fig:lrnpowerh03a}
\end{figure}

\begin{figure}[t]
    \centering
    \includegraphics[width=\columnwidth]{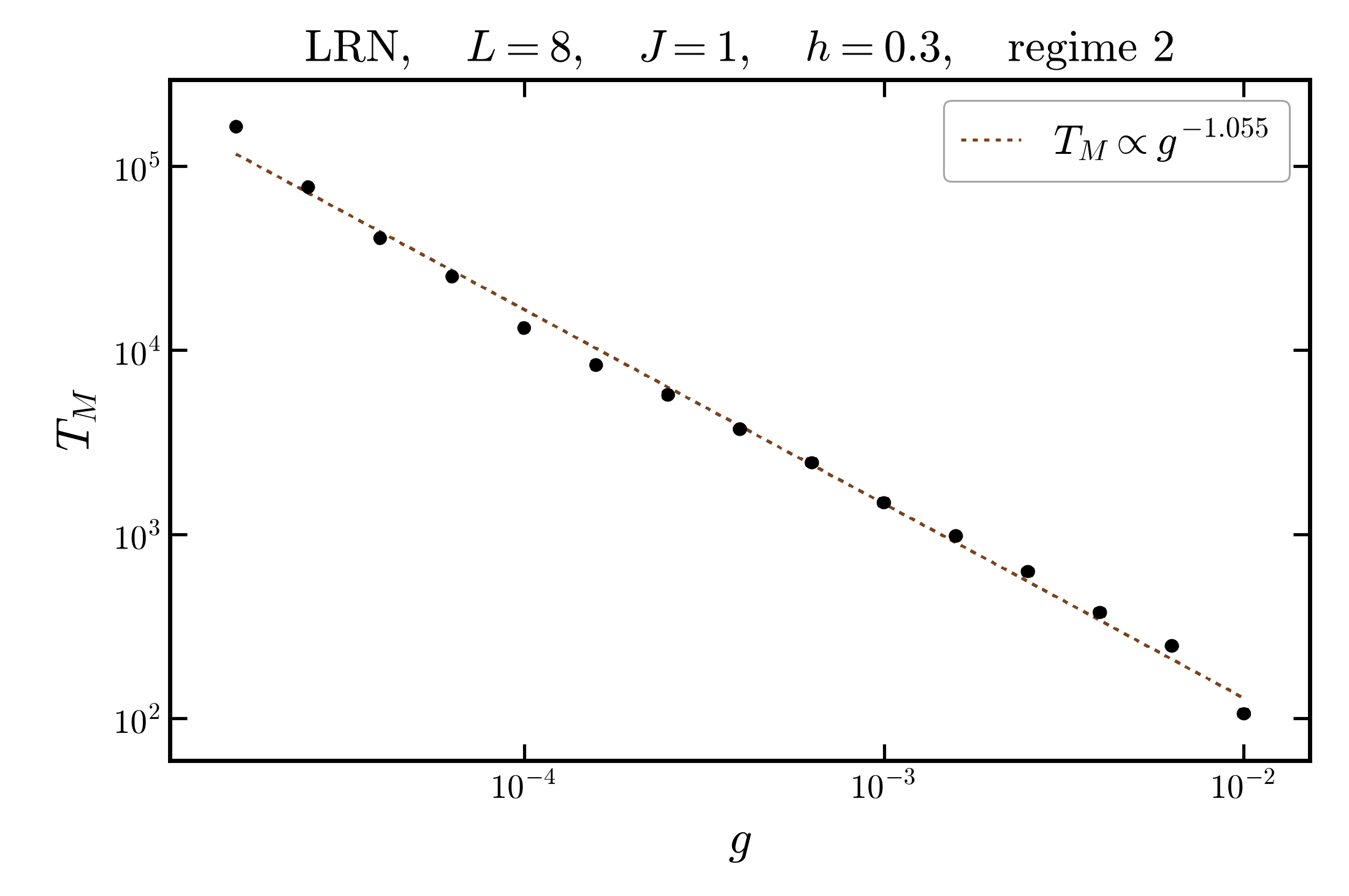}
    \caption{Second regime: melting time $T_M$ extracted from Fig.~\ref{fig:lrnentropyh03} for  $g\in[10^{-4.8},10^{-2}]$ .}
    \label{fig:lrnpowerh03b}
\end{figure}

The LRN regime therefore exhibits a more complex dependence on the model parameters than the SRN case.
Additional results for $h=1$ and system sizes $L=6,8,10,12,14$ are presented in Appendix~\ref{app:lrnscaling}.
The extracted melting-time exponents are close to $2$ for $L=6,8,12,14$, although the entropy thresholds and fitting intervals are not identical across system sizes.

A particularly interesting behavior is observed for $L=10$, where two distinct power-law scalings can be extracted from the same entropy dynamics, as shown in Figs.~\ref{fig:lrnapp_L10_entropy1}--%
\ref{fig:lrnapp_L10_fit2}.
Using the normalized entropy $S_{\mathrm{norm}}(t)=S(t)/[(L/2)\ln 2]$, the threshold $S_{\mathrm{norm}}=0.65$ yields $T_M\propto g^{-1.025}$, whereas the higher threshold $S_{\mathrm{norm}}=0.78$ yields $T_M\propto g^{-1.918}$.
The latter threshold probes a later stage of the entanglement growth and its power-law fit covers a more restricted range of perturbation strengths.

These results demonstrate that different stages of the same entanglement dynamics can exhibit substantially different melting-time exponents.
However, unlike the SRN regime at $g=2$, the LRN dynamics do not establish a comparably well-defined intermediate prethermal plateau.
The two fitted laws are therefore interpreted as threshold-dependent melting scalings, rather than as evidence of two distinct prethermal relaxation processes.

\section{Conclusion}
\label{sec:conclusion}

We studied the entanglement entropy dynamics of random product states under weak integrability breaking in the mixed-field Ising model, near its two integrable limits, which define the SRN and LRN regimes.
These two regimes relax in qualitatively different ways.
In the SRN regime away from $g = 2$, the departure from the integrable-state plateau follows $T_M \propto h^{-2}$, over the thresholds tested and for the parameters and system sizes examined.
At $g = 2$, the relaxation splits into two stages: the initial departure from integrable dynamics follows $T_M \propto h^{-1}$, while the subsequent melting of a well-defined intermediate prethermal plateau follows $T_{M_2} \propto h^{-2}$.
This provides evidence of multistage relaxation under a static Hamiltonian.
In the LRN regime, melting-time exponents close to both $1$ and $2$ also occur, but they shift with the model parameters, the system size, and the entropy threshold, without a comparably well-defined prethermal plateau. \barb{We present results for equal bipartition. However, we also considered all other possibilities of bipartition and found similar results.}
These observations have two consequences for the analysis of near-integrable dynamics.
First, an inverse-square scaling, often associated with Fermi's golden rule and with prethermal lifetimes, can arise without a distinct prethermal plateau.
Second, the scaling exponent is independent of the threshold only when the entropy curves collapse under a rescaling of time, as in the SRN regime away from $g = 2$; when this collapse fails, as in the LRN regime, different thresholds yield different exponents.
Melting-time exponents should therefore be read as measures of the departure from integrable dynamics, while prethermalization must be identified from the existence of a distinct intermediate plateau.
The microscopic origin of the two-stage relaxation at $g = 2$, and whether the threshold-dependent LRN scalings converge to a single exponent at longer times or larger system sizes, remain open questions for future work.

\section{Acknowledgments}

M.G. would like to acknowledge SECIHTI for funding his graduate studies and also the partial support from the project DGAPA-UNAM Project No. IN109523. BD acknowledges partial funding by the Deutsche Forschungsgemeinschaft (DFG, German Research Foundation) Project No. 290128388. BB acknowldeges support by the Luxembourg National Research Fund under Grant No.\ C24/MS/18940482/STAOpen.
We acknowledge the financial support from the Institute for Basic Science (IBS) in the Republic of Korea through Projects No. IBS-R024-D1, IBS-R041-D1-2026-a00, IBS-R041-D1-A3.

\bibliography{Bibliografy}

\appendix
\section{System-size and threshold dependence of LRN melting times}
\label{app:lrnscaling}

We present additional numerical results for the LRN regime at fixed transverse field $h=1$ and system sizes $L=6,8,10,12,14$.
Throughout this appendix, the entanglement entropy is normalized by its maximal value for an equal bipartition,
\begin{gather*} 
    S_{\mathrm{norm}}(t) = \frac{S(t)}{(L/2)\ln 2}.
\end{gather*}

The time evolution is shown for different strengths of the longitudinal perturbation $g$, with the color scale indicating $\log_{10}g$.
The blue reference curves correspond to the integrable limit $g=0$, while the horizontal threshold lines indicate the entropy values used to extract the melting times.
The black horizontal lines indicate the normalized Page values.

\begin{figure*}[htbp]
    \begin{minipage}{0.48\textwidth}
        \centering
        \includegraphics[width=\linewidth]{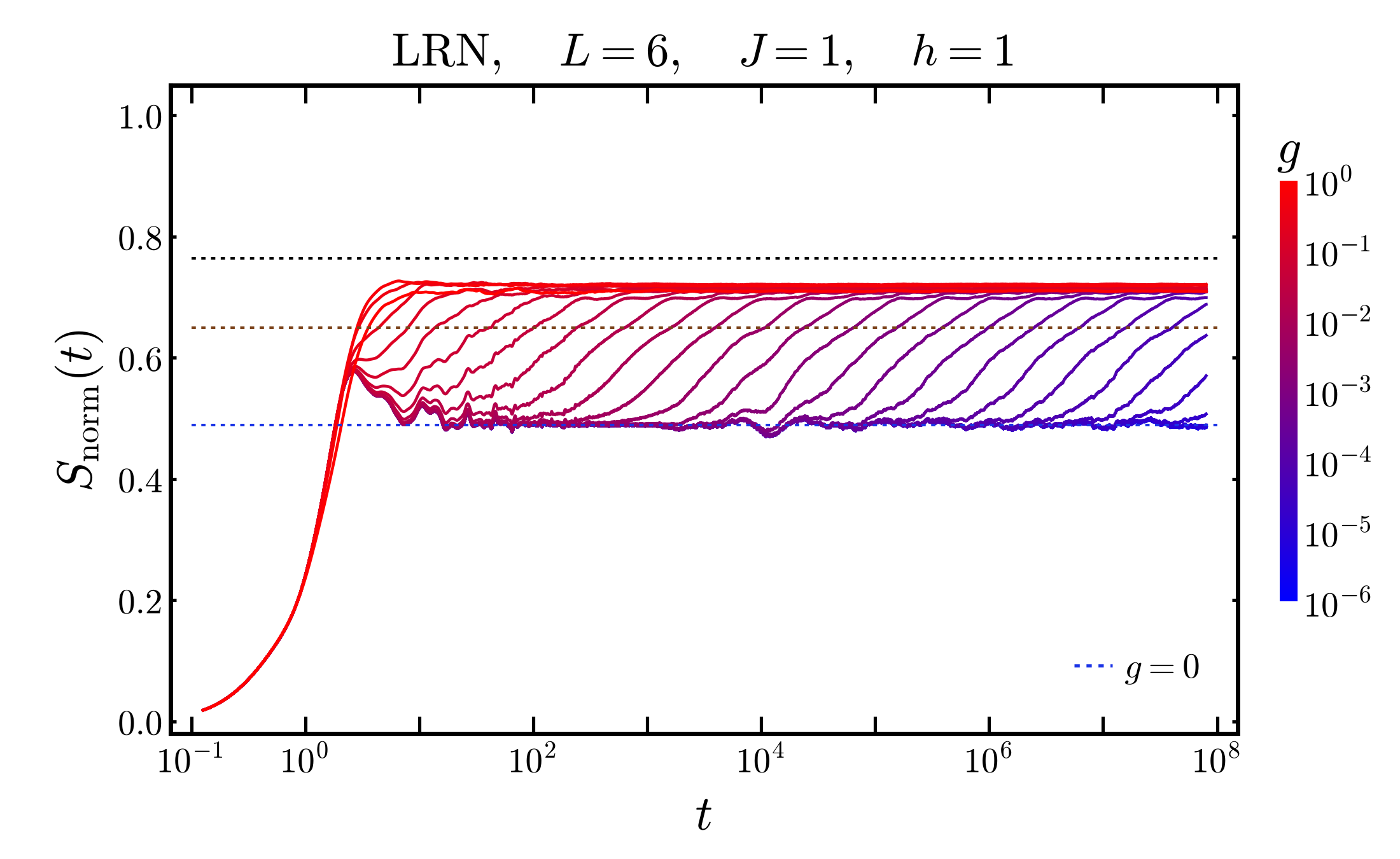}
        \caption{
            Normalized half-chain entanglement entropy $S_{\mathrm{norm}}(t)$ in the LRN regime for $L=6$, $J=1$ and $h=1$.
            The colored curves correspond to different longitudinal perturbations $g$, with smaller values shown in blue.
            The blue dotted line marks the integrable plateau ($g=0$), the brown dotted line marks the threshold $S_{\mathrm{norm}}=0.65$,
            and the black dotted line indicates the normalized Page value.
            The curves exhibit delayed entropy growth as the perturbation decreases.
        }
        \label{fig:lrnapp_L6_entropy}
    \end{minipage}\hfill
    \begin{minipage}{0.48\textwidth}
        \centering
        \includegraphics[width=\linewidth]{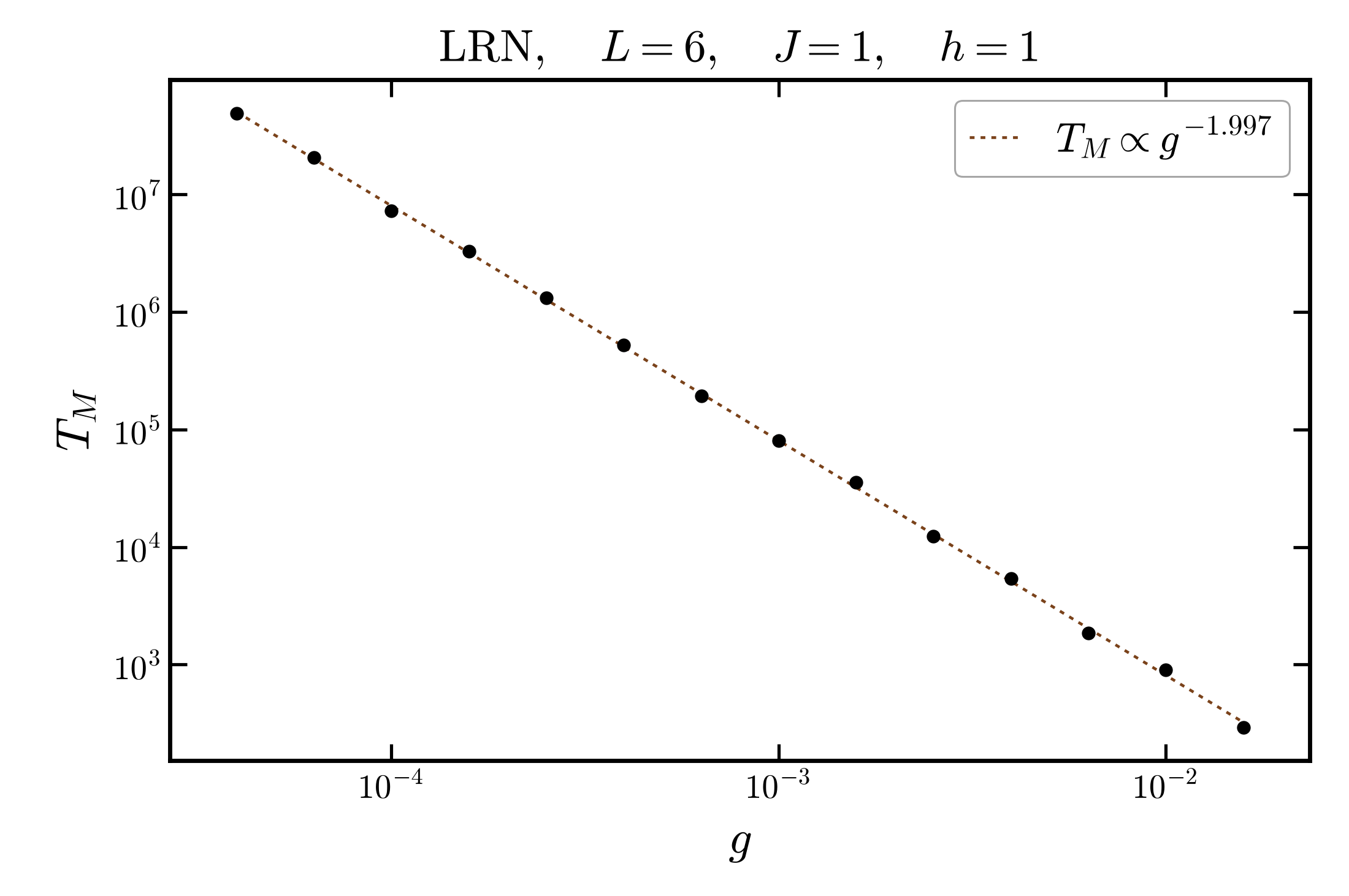}
        \caption{
            Melting time $T_M$ extracted from Fig.~\ref{fig:lrnapp_L6_entropy} using the threshold $S_{\mathrm{norm}}=0.65$.
            Black points denote the measured threshold-crossing times, and the brown dotted line is the power-law fit $T_M\propto g^{-1.997}$.
        }
        \label{fig:lrnapp_L6_fit}
    \end{minipage}
\end{figure*}

\begin{figure*}[htbp]
    \begin{minipage}{0.48\textwidth}
        \centering
        \includegraphics[width=\linewidth]{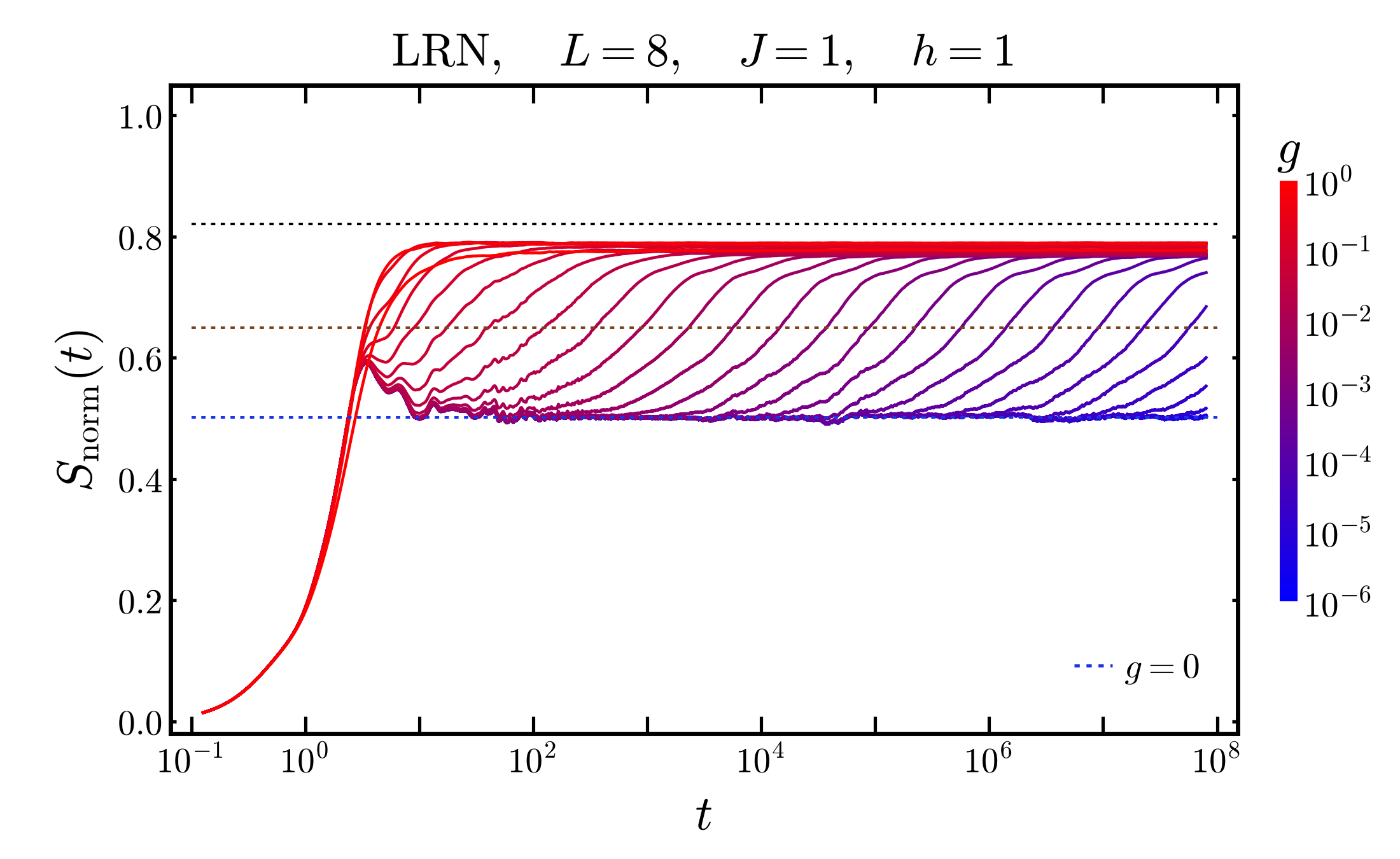}
        \caption{
            Normalized half-chain entanglement entropy in the LRN regime for $L=8$, $J=1$ and $h=1$.
            The colored curves show the evolution for different values of the integrability-breaking perturbation $g$.
            The blue dotted line marks the integrable plateau ($g=0$), the brown dotted line marks $S_{\mathrm{norm}}=0.65$,
            and the black dotted line indicates the normalized Page value.
            The delayed growth becomes increasingly pronounced for weaker perturbations.
        }
        \label{fig:lrnapp_L8_entropy}
    \end{minipage}\hfill
    \begin{minipage}{0.48\textwidth}
        \centering
        \includegraphics[width=\linewidth]{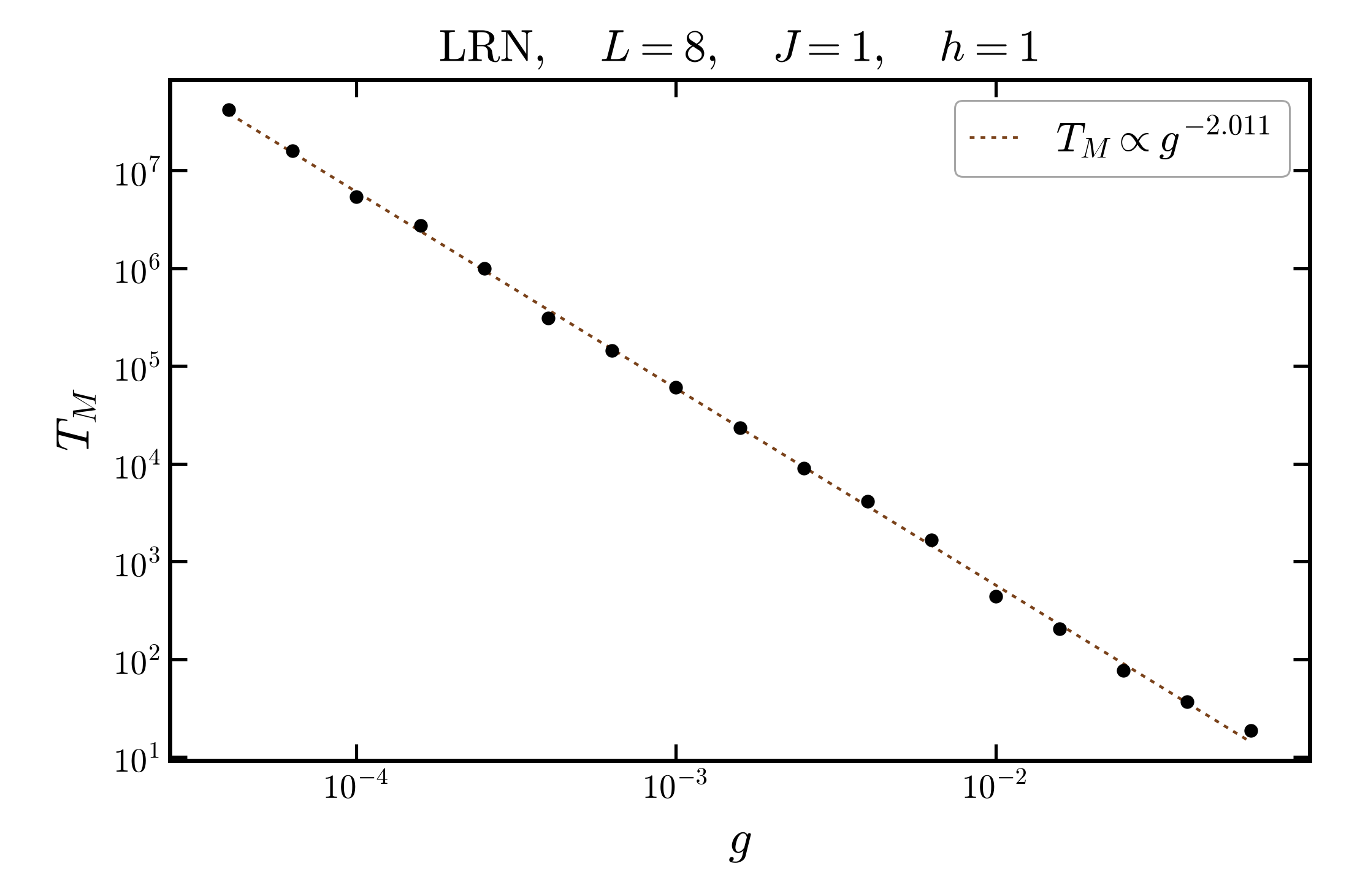}
        \caption{
            Melting time $T_M$ extracted from Fig.~\ref{fig:lrnapp_L8_entropy} using $S_{\mathrm{norm}}=0.65$.
            The black points represent the numerical results and the brown dotted line shows the power-law fit $T_M\propto g^{-2.011}$.
        }
        \label{fig:lrnapp_L8_fit}
    \end{minipage}
\end{figure*}

\begin{figure*}[htbp]
    \begin{minipage}{0.48\textwidth}
        \centering
        \includegraphics[width=\linewidth]{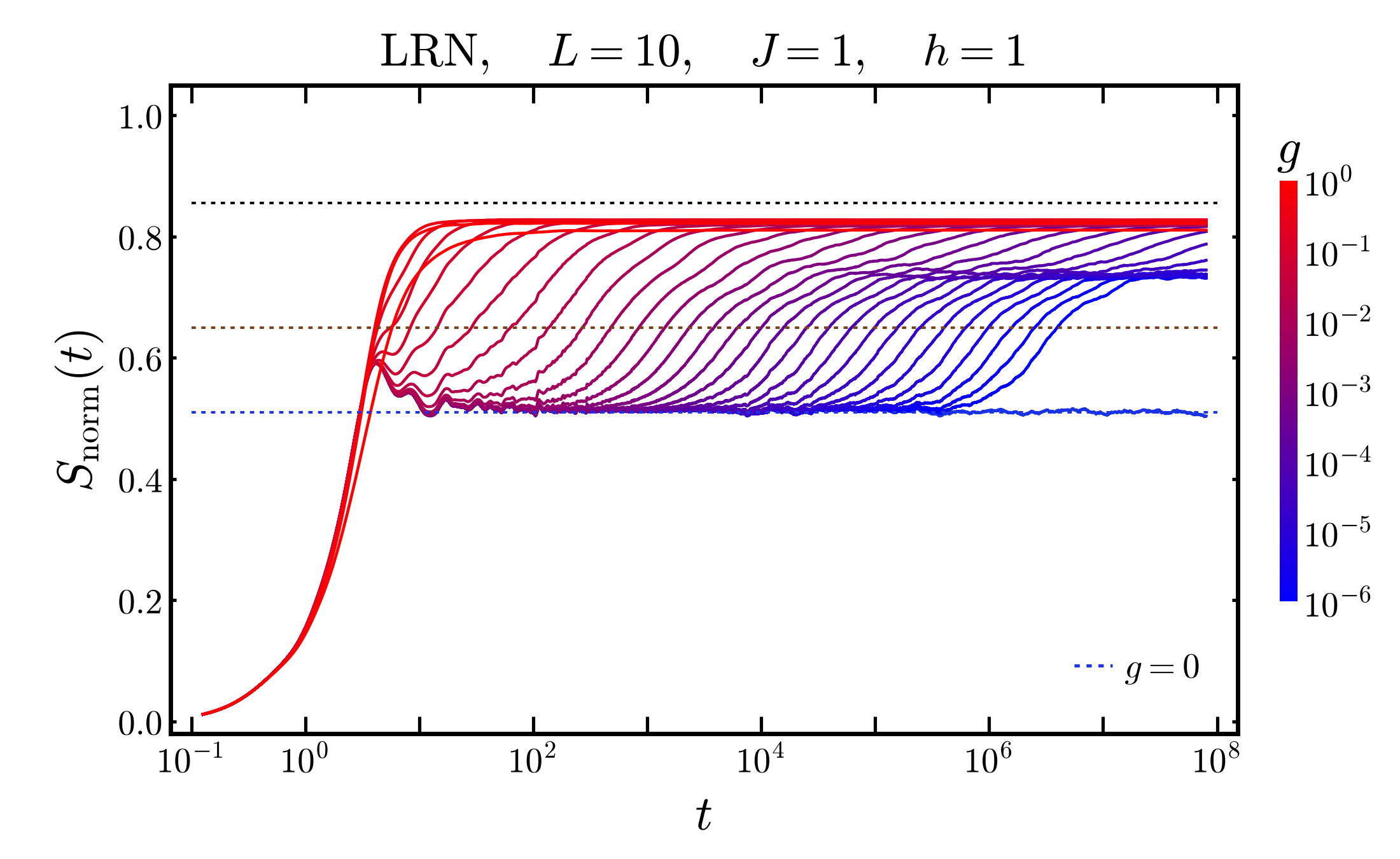}
        \caption{
            Normalized half-chain entanglement entropy for the LRN regime with $L=10$, $J=1$ and $h=1$.
            The colored curves correspond to different integrability-breaking perturbations $g$.
            After an initial growth, the entropy remains close to the integrable plateau (blue dotted line) before undergoing a delayed increase.
            The brown dotted line at $S_{\mathrm{norm}}=0.65$ defines the lower threshold used to characterize this departure from integrable dynamics.
            The black dotted line indicates the normalized Page value.
        }
        \label{fig:lrnapp_L10_entropy1}
    \end{minipage}\hfill
    \begin{minipage}{0.48\textwidth}
        \centering
        \includegraphics[width=\linewidth]{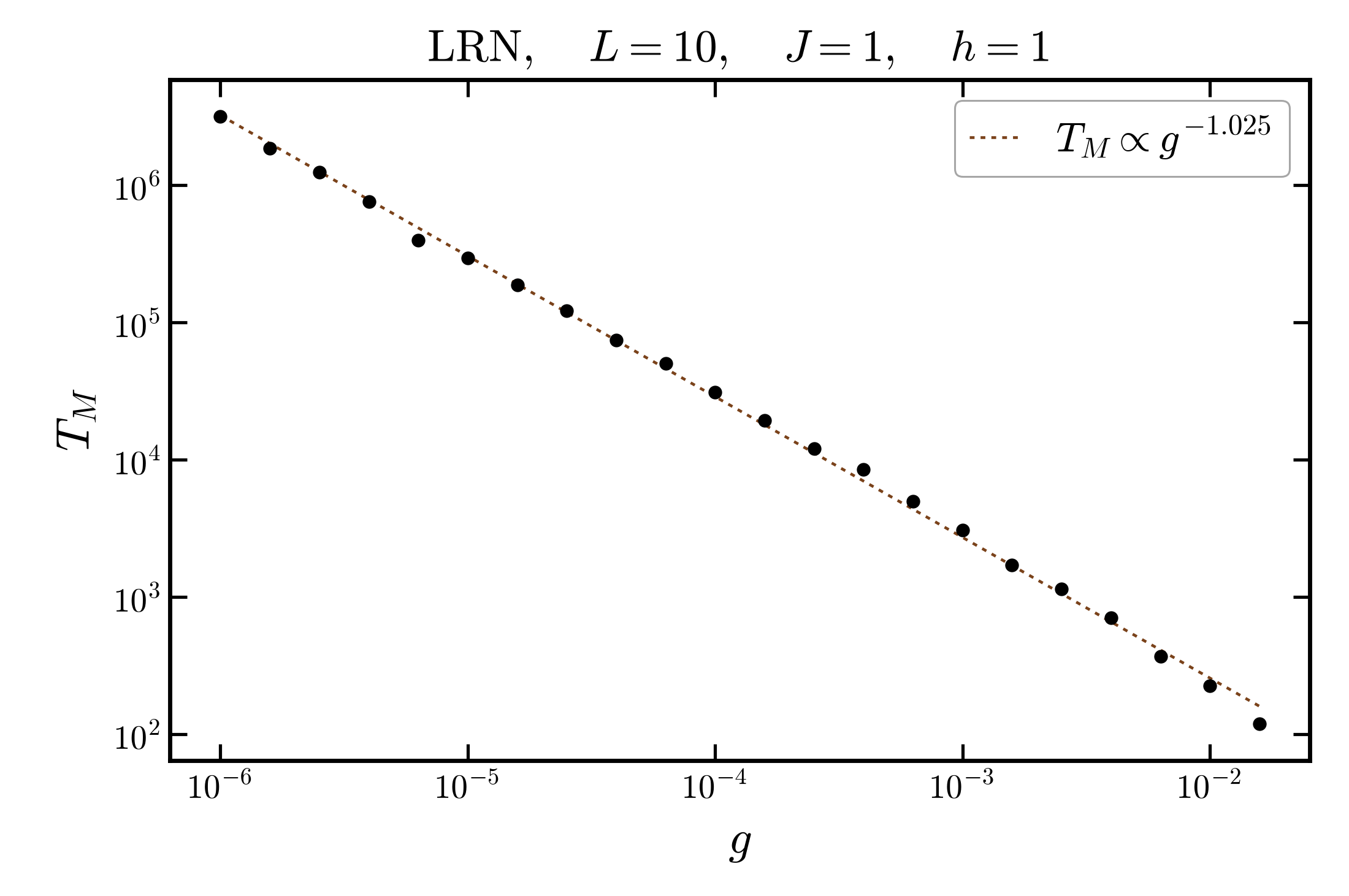}
        \caption{
            Melting time extracted from Fig.~\ref{fig:lrnapp_L10_entropy1} using the normalized entropy threshold $S_{\mathrm{norm}}=0.65$.
            The numerical threshold-crossing times (black points) follow $T_M\propto g^{-1.025}$, indicated by the brown dotted fit.
            This exponent characterizes the earlier stage of the delayed entropy growth.
        }
        \label{fig:lrnapp_L10_fit1}
    \end{minipage}
\end{figure*}

\begin{figure*}[htbp]
    \begin{minipage}{0.48\textwidth}
        \centering
        \includegraphics[width=\linewidth]{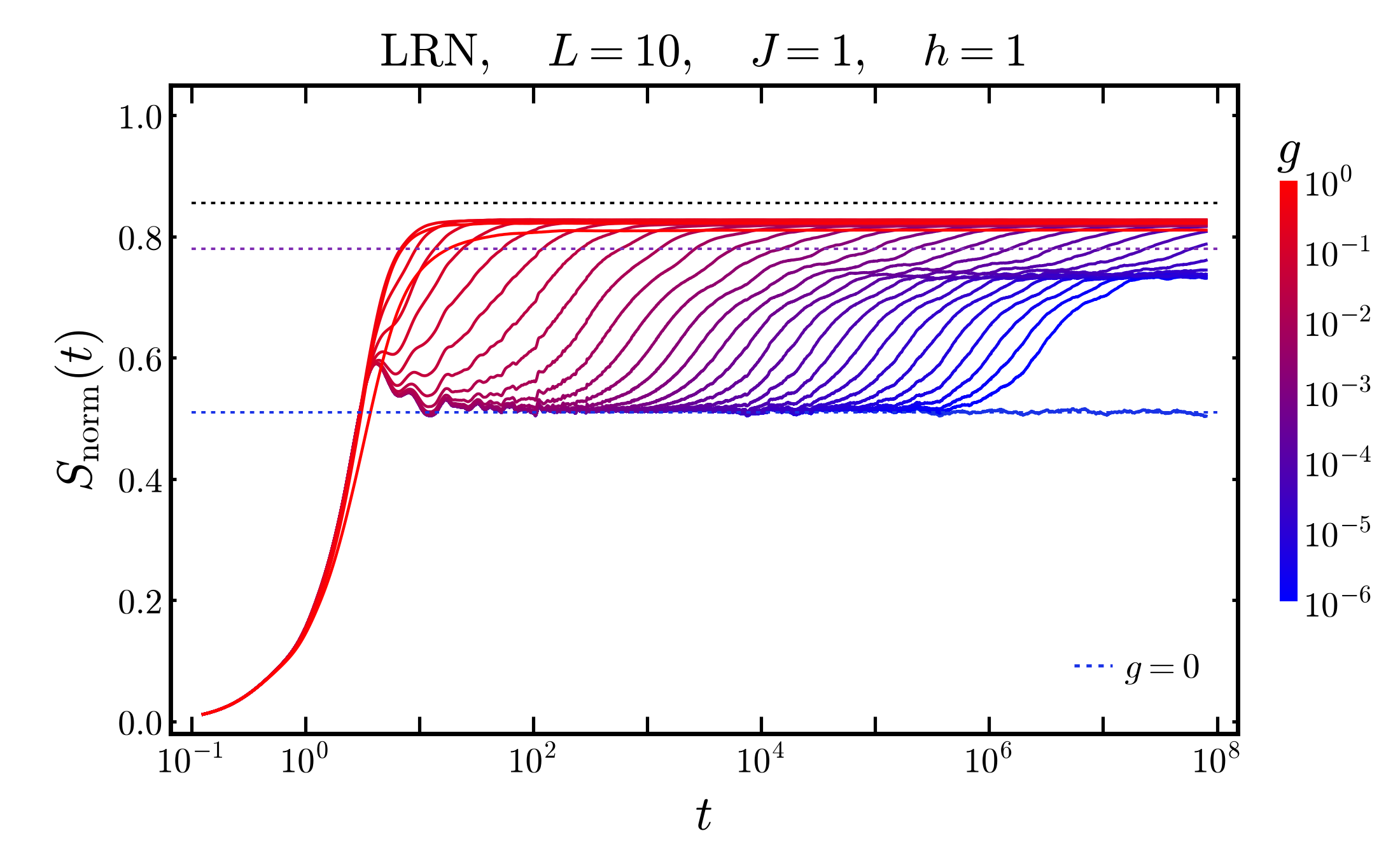}
        \caption{
            Same normalized entanglement entropy dynamics as in Fig.~\ref{fig:lrnapp_L10_entropy1} for $L=10$, $J=1$ and $h=1$,
            now analyzed using the higher threshold $S_{\mathrm{norm}}=0.78$, indicated by the purple dotted line.
            This threshold probes a later stage of the delayed entropy growth, closer to the upper saturation region.
            The blue dotted line marks the integrable plateau ($g=0$), while the black dotted line indicates the normalized Page value.
        }
        \label{fig:lrnapp_L10_entropy2}
    \end{minipage}\hfill
    \begin{minipage}{0.48\textwidth}
        \centering
        \includegraphics[width=\linewidth]{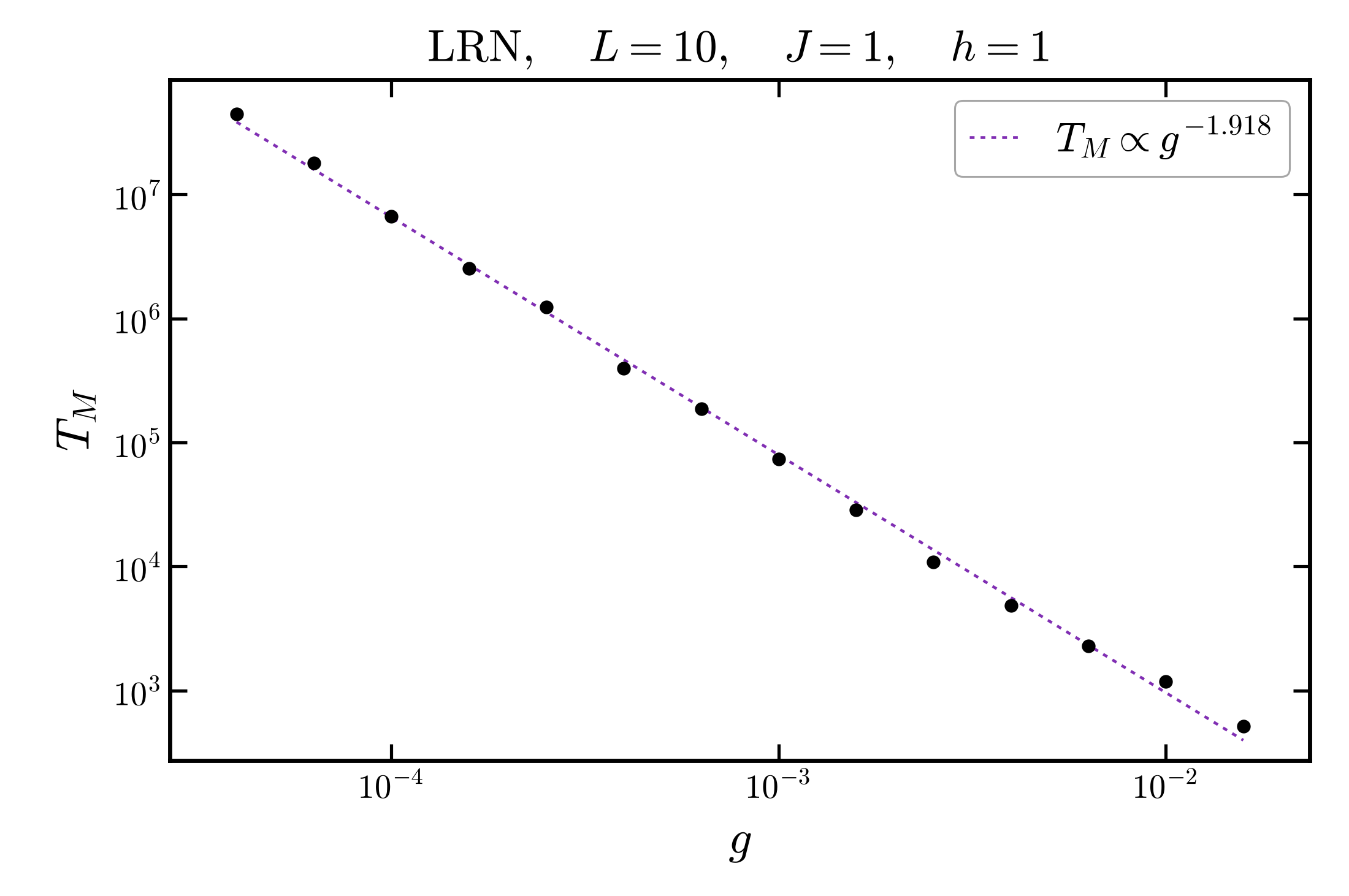}
        \caption{
            Melting time extracted from Fig.~\ref{fig:lrnapp_L10_entropy2} using the higher normalized entropy threshold $S_{\mathrm{norm}}=0.78$.
            The fit starts at $g=10^{-4.4}\approx4.0\times10^{-5}$.
            Black points denote the measured threshold-crossing times, and the purple dotted line gives $T_M\propto g^{-1.918}$.
            The exponent is close to $2$, in contrast to the exponent $1.025$ obtained at the lower threshold in Fig.~\ref{fig:lrnapp_L10_fit1}.
            The two fits probe different stages of the same entropy evolution and cover different perturbation intervals.
        }
        \label{fig:lrnapp_L10_fit2}
    \end{minipage}
\end{figure*}

Figures~\ref{fig:lrnapp_L6_entropy}--\ref{fig:lrnapp_L8_fit} show the results for $L=6$ and $L=8$.
The melting times extracted using $S_{\mathrm{norm}}=0.65$ follow $T_M\propto g^{-1.997}$ and $T_M\propto g^{-2.011}$, respectively. 
For $L=10$, the entropy dynamics exhibit a pronounced nonuniform growth, allowing different stages of the relaxation to be characterized using distinct entropy thresholds.
Figures~\ref{fig:lrnapp_L10_entropy1} and~\ref{fig:lrnapp_L10_fit1} show the first measurement, obtained using $S_{\mathrm{norm}}=0.65$.
The corresponding melting time follows $T_M\propto g^{-1.025}$.
The same entropy dynamics are shown in Fig.~\ref{fig:lrnapp_L10_entropy2}, now using the higher threshold
$S_{\mathrm{norm}}=0.78$.
The corresponding melting time, shown in Fig.~\ref{fig:lrnapp_L10_fit2}, follows $T_M\propto g^{-1.918}$.
The higher threshold probes a later stage of the entropy growth and can be reached within the numerical time window over a more restricted range of perturbation strengths.
The change from an exponent close to $1$ to one close to $2$ is therefore obtained without changing the Hamiltonian parameters, system size, or initial-state ensemble, but by probing different entropy thresholds.
These measurements do not establish that both exponents characterize the same asymptotic regime as $g\to 0$. 
The $L=10$ comparison is particularly instructive because it separates the dependence on the entropy threshold from changes in the microscopic model.
The two fits are obtained for identical $L$, $J$, $h$, and initial-state ensemble.
Their different exponents therefore reflect the choice of the entropy level used to define the characteristic time, together with the different perturbation intervals accessible to each threshold.
This observation reinforces the need to distinguish a threshold-dependent relaxation time from the lifetime of a physically identified prethermal plateau.

\begin{figure*}[htbp]
    \begin{minipage}{0.48\textwidth}
        \centering
        \includegraphics[width=\linewidth]{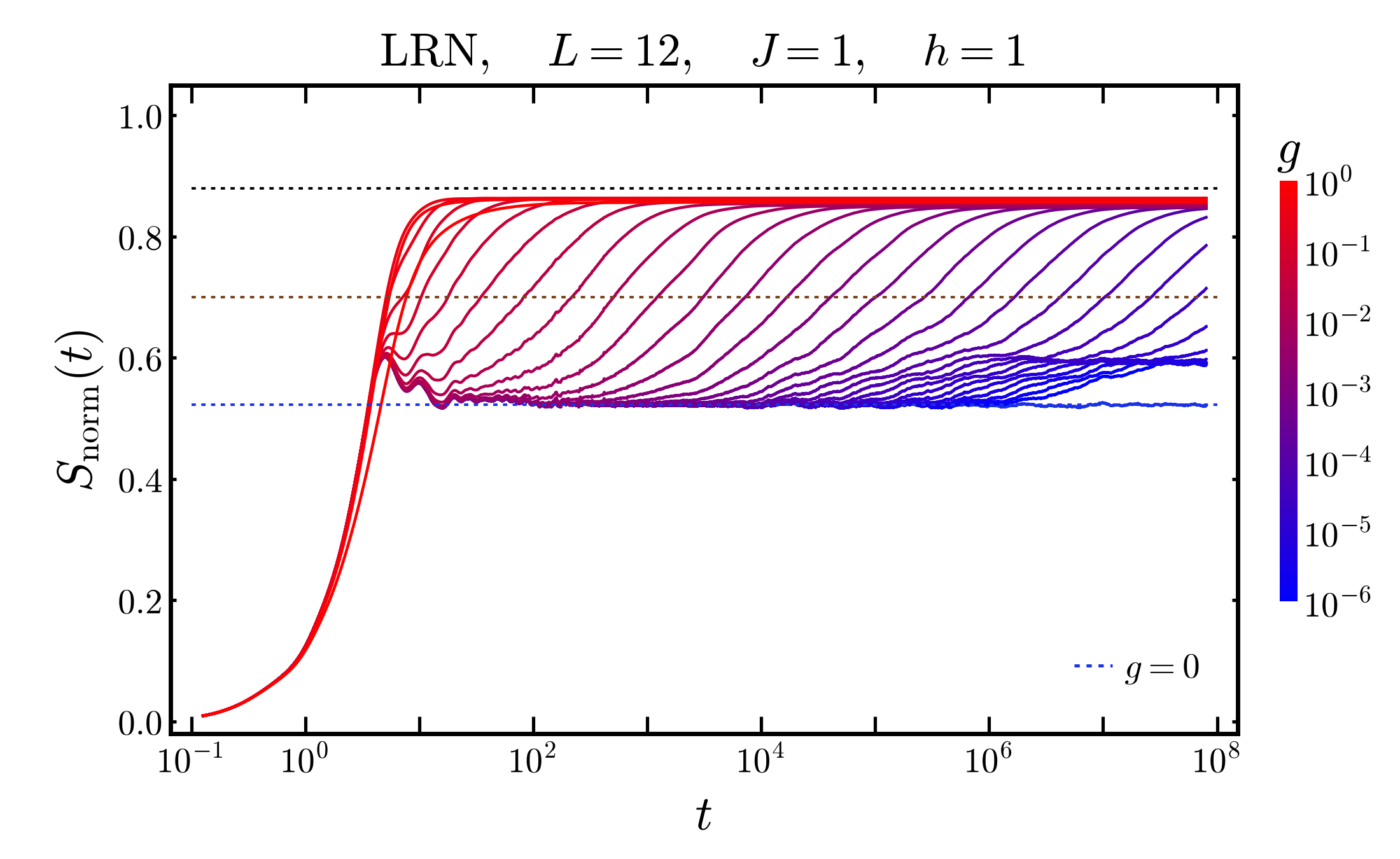}
        \caption{
            Normalized half-chain entanglement entropy in the LRN regime for $L=12$, $J=1$ and $h=1$.
            The colored curves correspond to different values of $g$.
            The blue dotted line marks the integrable plateau ($g=0$), while the black dotted line denotes the normalized Page value.
            The brown dotted line marks the melting-time extraction threshold at $S_{\mathrm{norm}}=0.70$.
            The delayed entropy growth extends to progressively longer times as $g$ decreases.
        }
        \label{fig:lrnapp_L12_entropy}
    \end{minipage}\hfill
    \begin{minipage}{0.48\textwidth}
        \centering
        \includegraphics[width=\linewidth]{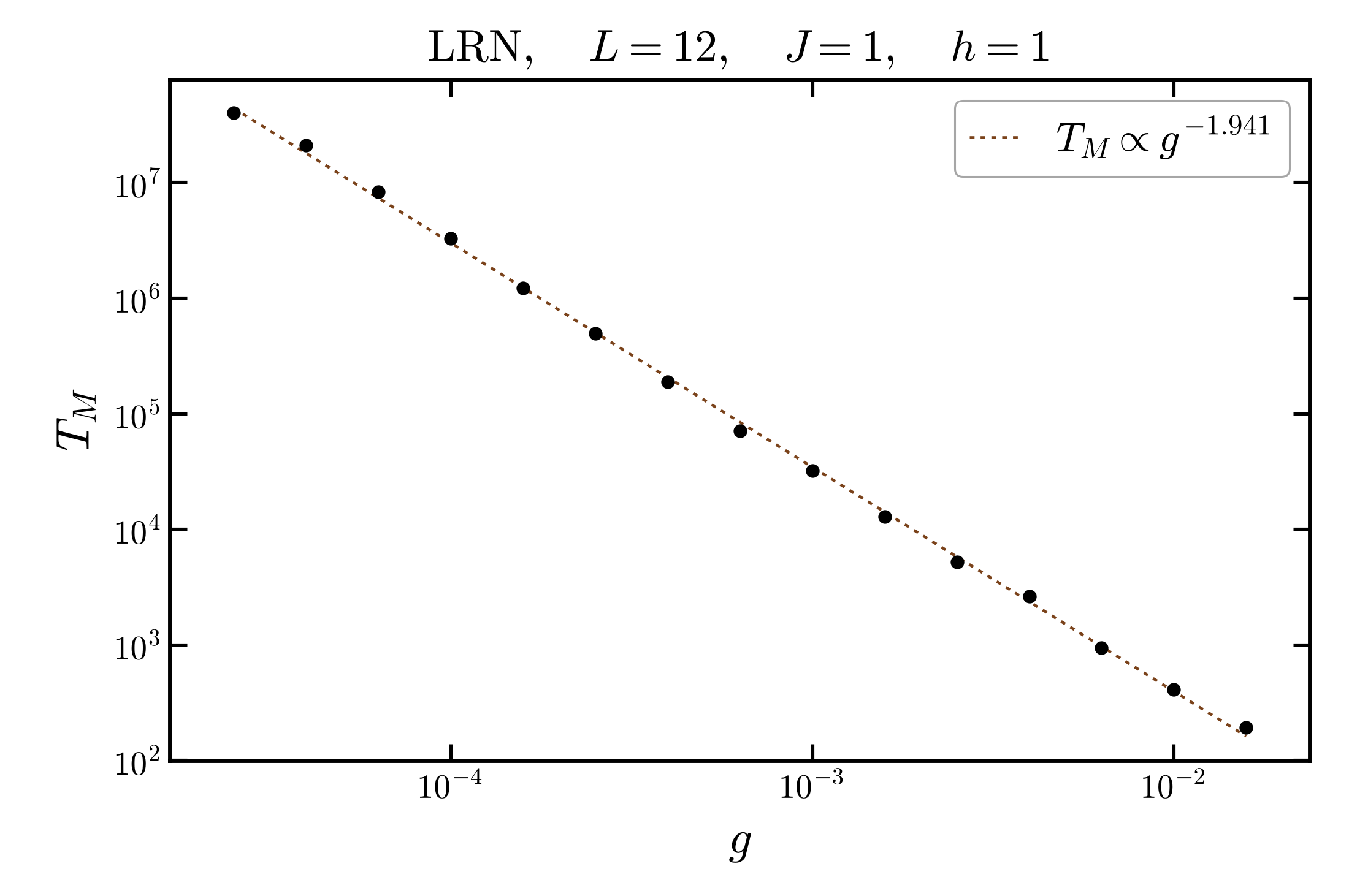}
        \caption{
            Melting time $T_M$ extracted from the threshold shown in Fig.~\ref{fig:lrnapp_L12_entropy}.
            Black points denote the numerical results and the brown dotted line indicates the fitted scaling $T_M\propto g^{-1.941}$.
        }
        \label{fig:lrnapp_L12_fit}
    \end{minipage}
\end{figure*}

\begin{figure*}[htbp]
    \begin{minipage}{0.48\textwidth}
        \centering
        \includegraphics[width=\linewidth]{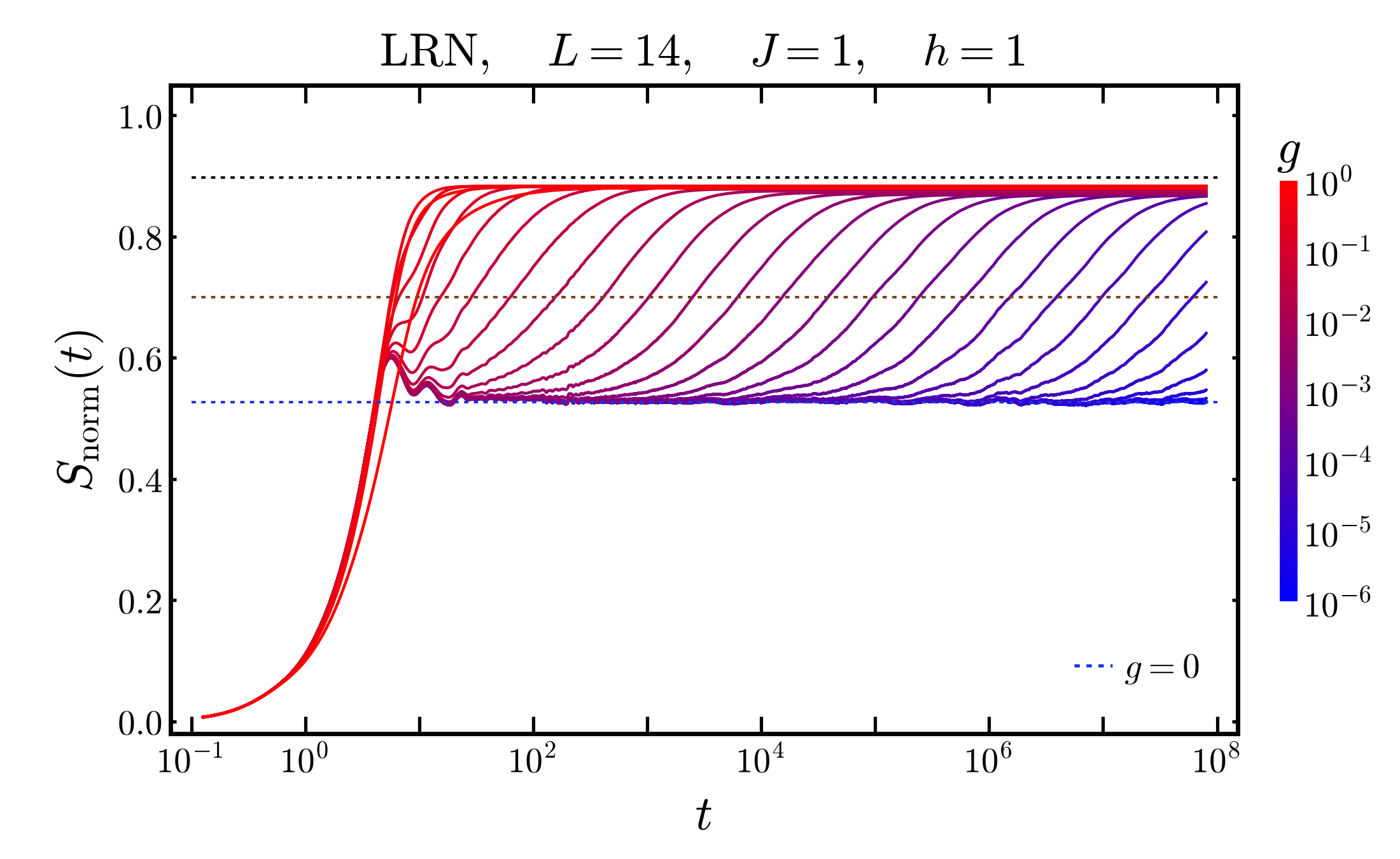}
        \caption{
            Normalized half-chain entanglement entropy in the LRN regime for $L=14$, $J=1$ and $h=1$.
            The colored curves show the evolution for different perturbation strengths $g$, with smaller values in blue.
            The blue dotted line marks the integrable plateau, while the black dotted line denotes the normalized Page value.
            The brown dotted line marks the melting-time extraction threshold at $S_{\mathrm{norm}}=0.70$.
        }
        \label{fig:lrnapp_L14_entropy}
    \end{minipage}\hfill
    \begin{minipage}{0.48\textwidth}
        \centering
        \includegraphics[width=\linewidth]{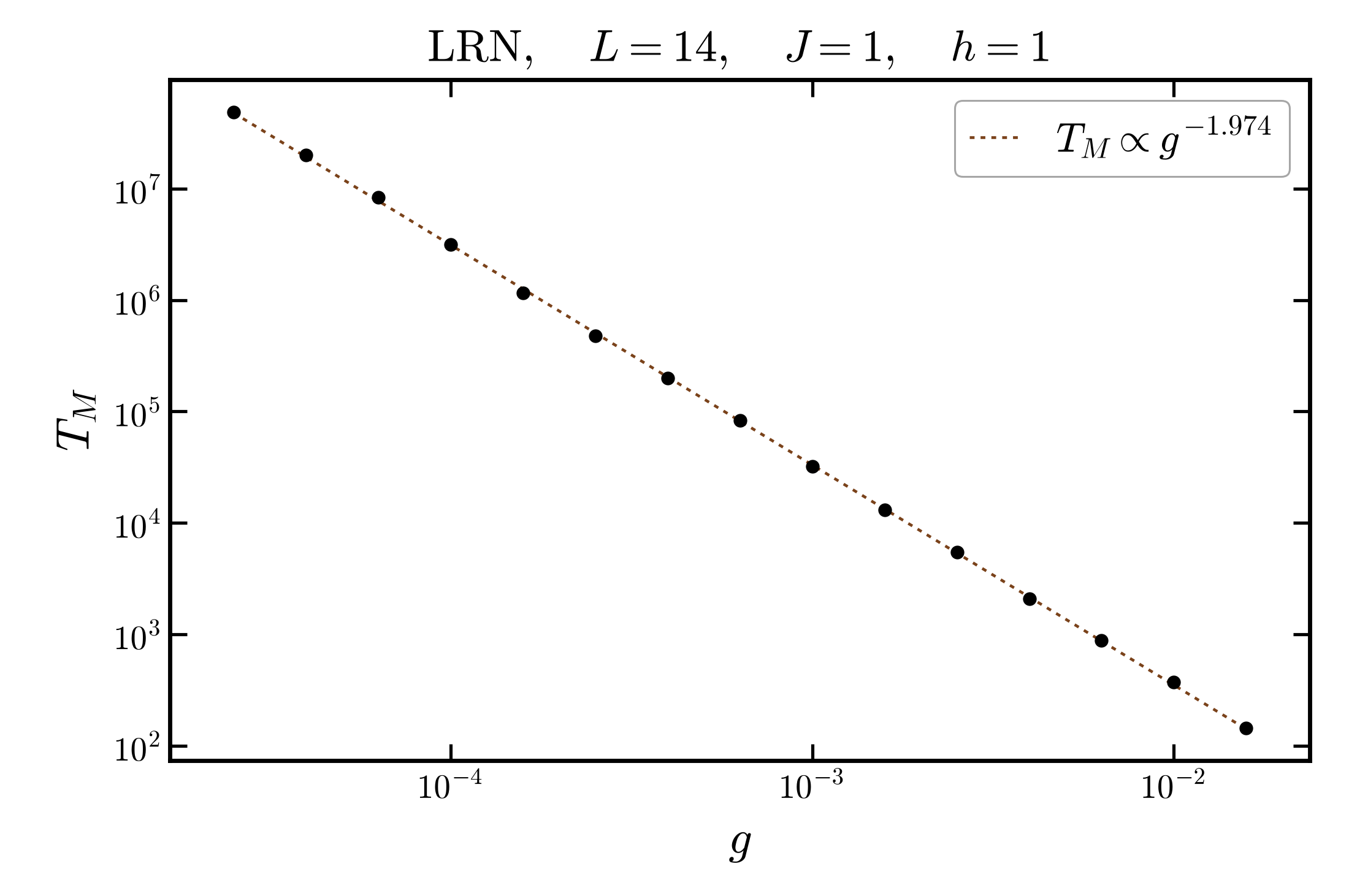}
        \caption{
            Melting time $T_M$ extracted from Fig.~\ref{fig:lrnapp_L14_entropy}.
            Black points are the numerical threshold-crossing times, and the brown dotted line represents the fit $T_M\propto g^{-1.974}$.
            The exponent is again close to $2$ over the displayed fitting interval.
        }
        \label{fig:lrnapp_L14_fit}
    \end{minipage}
\end{figure*}

Finally, Figs.~\ref{fig:lrnapp_L12_entropy}--\ref{fig:lrnapp_L14_fit} show the results for $L=12$ and $L=14$, with fitted exponents $1.941$ and $1.974$, respectively.
Although several system sizes yield exponents close to $2$, the threshold dependence observed at $L=10$ prevents a straightforward interpretation of these data as a universal system-size scaling law.
These results further illustrate the complexity of LRN entanglement relaxation.
In particular, the existence of different melting-time exponents does not by itself imply the formation and subsequent melting of a distinct prethermal plateau.

\end{document}